\documentclass[%
twocolumn,secnumarabic,amssymb, amsmath, tightenlines,nobibnotes, aps, prl,longbibliography]{revtex4-1}

\usepackage[pdftex,colorlinks=true,pdfstartview=Fit]{hyperref}		

\let\oldmarginpar\marginpar
\renewcommand\marginpar[1]{\-\oldmarginpar[\raggedleft\footnotesize #1]%
	{\raggedright\footnotesize #1}}

\usepackage{color}			

\usepackage{bm}		
\usepackage{amssymb,amsmath}
\usepackage{graphicx}

\usepackage{pdfpages}

\begin{document}

	\title{Erasure without work in an asymmetric, double-well potential}

	\author{Mom\v cilo Gavrilov}
	\email[email: ]{momcilog@sfu.ca}
	
	\author{John Bechhoefer}
	\email[email: ]{johnb@sfu.ca}
	
	\affiliation{Department of Physics, Simon Fraser University, Burnaby, British Columbia, V5A 1S6, Canada}
	
	\begin{abstract}
		According to Landauer's principle, erasing a memory requires an average work of at least $kT\ln2$ per bit.  Recent experiments have confirmed this prediction for a one-bit memory represented by a symmetric double-well potential.  Here, we present an experimental study of erasure for a memory encoded in an asymmetric double-well potential.  Using a feedback trap, we find that the average work to erase can be less than $kT\ln2$.  Surprisingly, erasure protocols that differ subtly  give measurably different values for the asymptotic work, a result we explain by showing that one protocol is symmetric with the respect to time reversal, while the other is not.  The differences between the protocols help clarify the distinctions between thermodynamic and logical reversibility.
	\end{abstract}
	
	\maketitle
	\noindent \textit{Introduction}.  Landauer's principle states that erasing a one-bit memory requires an average work of at least $kT\ln2$ \cite{landauer61,sagawa14}, with the lower bound achieved in the quasistatic limit.  It plays a key role in sharpening our understanding of the second law of thermodynamics and of the interplay between information and thermodynamics, an issue first raised Maxwell \cite{maxwell1871}, developed in important contributions by Szilard \cite{szilard29} and Bennett \cite{bennett82} but also subject to a long, sometimes confused discussion \cite{leff03}.  Recent experiments have confirmed Landauer's prediction in simple systems:  in a one-bit memory represented by a symmetric double-well potential \cite{berut12,Jun14}, in memory encoded by nanomagnetic bits \cite{Hong16,Martini16}, and even in quantum bits \cite{Peterson16}.  These successes have helped to create an extended version of stochastic thermodynamics \cite{sekimoto10,seifert12} that views information as another kind of thermodynamic resource, on the same footing as heat, chemical energy, and other sources of work \cite{parrondo15}.  This new way of looking at thermodynamics has led to experimental realizations of \textit{information engines} (``Maxwell demons") \cite{toyabe10,koski14a,koski15,camati16}.
	
	Despite its success in simple situations, Landauer's principle remains untested in more complex cases, such as systems where the symmetry between states is broken.  This case, briefly mentioned but not pursued in Landauer's original paper \cite{landauer61}, was followed up in later theoretical work \cite{shizume95,fahn96,barkeshli05,Sagawa09,turgut09,KlagesBook13,sagawa14,boyd16}.  
	In addition, erasure in asymmetric states can be interpreted as situations where the initial system is out of global thermodynamic equilibrium.  Since nonequilibrium settings are ubiquitous in biological systems, it is important to establish basic scenarios in simpler settings to understand more clearly, for example, why common biological systems may not be able to reach ultimate thermodynamic limits \cite{ouldridge16}.
	
	
	Here, we explore the broken-symmetry case experimentally by studying erasure in a memory represented by an \textit{asymmetric}, double-well potential.  While we find broad agreement with the main predictions of theoretical work, we also find that surprising subtleties distinguish nominally similar protocols.
	
	
	In this paper, we follow the analysis of Sagawa and Ueda, who argue that the average work to erase a one-bit memory can be less than $kT\ln 2$, if the volume in phase space corresponding to each state is different \cite{Sagawa09,sagawa14}.  Indeed, they predict that the work can even be negative, when the asymmetry (volume ratio) is sufficiently high.  In all cases, however, the $kT \ln 2$-per-bit bound should still hold over the entire cycle of measurement and erasure.  In a Comment, Dillenschneider and Lutz linked the reduction of  required average work from $kT \ln 2$ to the fact that the proposed erasure cycle starts in a nonequilibrium state, in contrast to the classical case where it is assumed that erasure starts from equilibrium \cite{Dill10}.
	The proposed explanation lies in a proper accounting for different number of microstates associated with each \textit{information-bearing} degree of freedom in the system; however, the interpretation of the Gibbs-Shannon entropy for the asymmetric state is still subject to debate \cite{Sagawa10}.  Here, we investigate experimentally these questions.
	
	Following the scheme of Sagawa and Ueda, our one-bit memory is realized using an asymmetric double-well potential.  We will see that the average work to erase an asymmetric bit can be below $kT\ln2$ when the erasure cycle is performed arbitrarily slowly.  For high-enough asymmetry, this work can vanish, or even be negative.  We also find, perhaps surprisingly, that not all erasure protocols can achieve the expected limits, even when extrapolated to very slow cycle times.
	
	\noindent \textit{Feedback trap.} Our classical one-bit memory consists of an overdamped silica bead of diameter 1.5~$\mu$m trapped in a virtual time-dependent double-well potential imposed by a feedback (or ABEL) trap \cite{cohen05b,cohen05d,jun12,gavrilov13, Gavrilov16a}. A feedback trap periodically measures particle position and, after each measurement, applies a force imposed by the potential $U(x, y, t)$ \cite{jun12, gavrilov14, gavrilov15, Gavrilov16a}.  Feedback traps have been used to measure properties of particles and molecules \cite{cohen06a,cohen05d,wang14,cohen07a,Goldsmith2010,Fields2011,germann14,kayci14} and to explore fundamental questions in the non-equilibrium statistical mechanics of small systems \cite{cohen05b,cohen05d,jun12,gavrilov13,Jun14,lee15,gavrilov15}.
	Since there is no physical potential that traps the particle---only the approximation imposed by the rapid feedback loop---the potential is \textit{virtual}.  
	
	%
	
	\noindent\textit{Virtual potential.} The virtual double-well potential $U(x, y, t)$ is parametrized for independent control of the barrier, asymmetry, and tilt.  This is straightforward to control using a virtual potential, but very difficult using an ordinary, physical potential.  Combining those operations, we implement the asymmetric erasure protocol \cite{Sagawa09}.  
	
\begin{figure}[tb]
		\begin{center}
			\includegraphics[width=8.6cm]{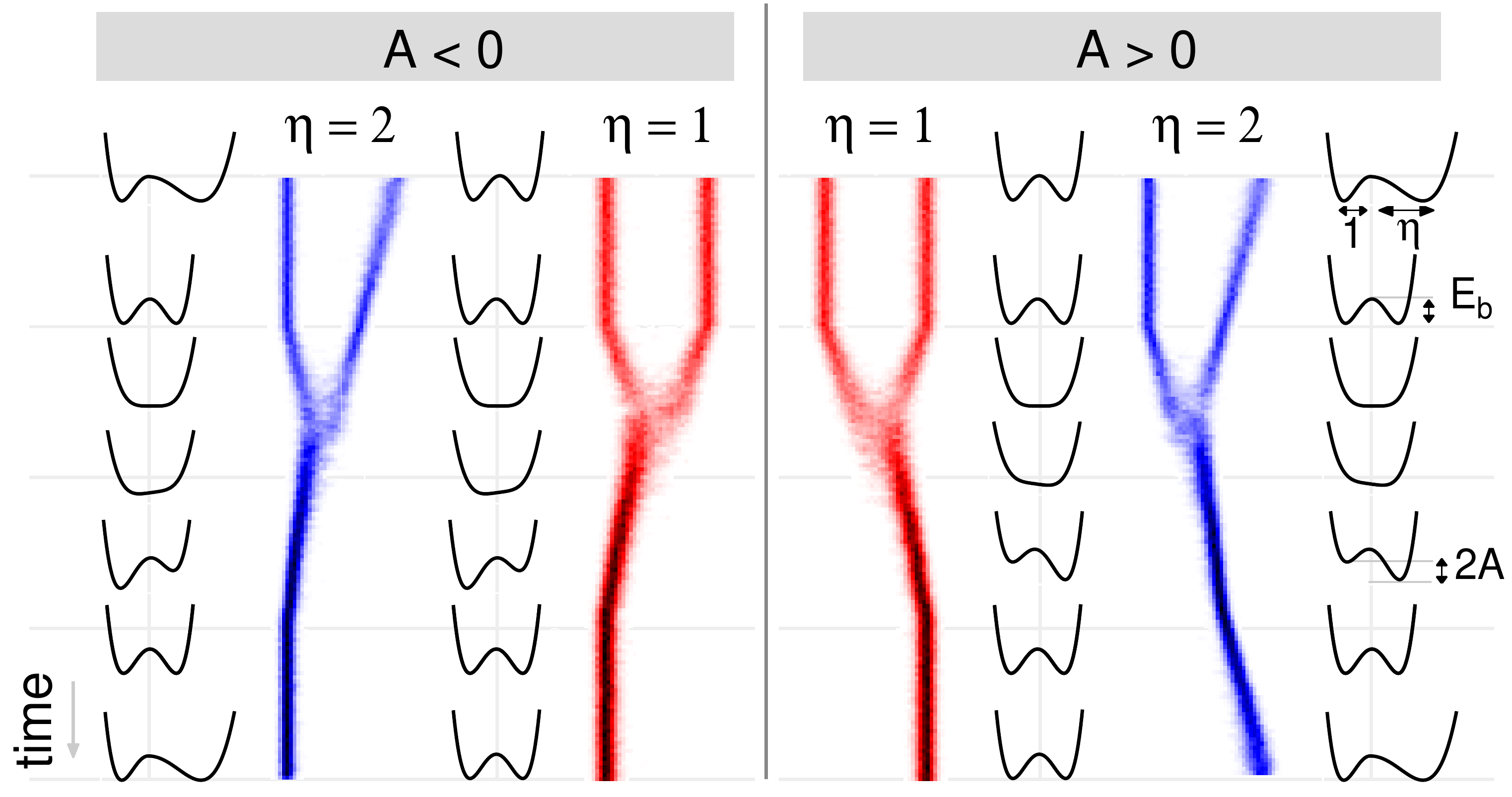}
		\end{center}
		\caption { \label{fig:trajectories} 
			(color online) Erasure protocols and time evolution of probability for symmetric ($\eta=1$) and asymmetric ($\eta=2$) potentials.   The one-bit memory is erased to the left ($A<0$) and right ($A>0$) wells.  The images are two-dimensional histograms, with intensity $\propto P(x,t)$ the occupation probability for a particle in a time-dependent, double-well potential.  The separation between barrier maximum and left well minimum is $x_0 =  0.77~\mu$m.
		}
	\end{figure}

	The ratio between the number of microstates in the two wells is controlled by the asymmetry parameter $\eta \geq 1$.  For $\eta=1$ the potential is symmetric and for $\eta>1$ the right-hand well is larger (see Fig.~\ref{fig:trajectories}).  The tilt sign determines the well used for erasure, with the tilt scale set by $|A| / kT = 0.25$.  See \cite{Note1} for the explicit functional form of the potential and control functions for barrier height, tilt, and asymmetry.
	
	\noindent\textit{Data acquisition.} For each measurement, we choose an asymmetry factor $\eta$, a tilt direction (sign), and an erasure cycle time $\tau$.  We assign an equal probability for a particle to be in the left or the right well.  The initial condition is prepared by applying a strong harmonic trap for $0.5$ s centered on $-x_0$ for half the measurements and on $+\eta x_0$ for the other half.   After the initial condition has been set, we switch abruptly to an asymmetric double well and then let the particle relax for $1$ s.  
	
	Although our protocol creates an equal number of left and right initial states, the ensemble is not in global equilibrium,  because both states have the same occupation probability ($p_i=0.5$) but different spatial extents.  
	
	At the beginning and end of each erasure cycle, the barrier is set to $E_b/kT=13$, which prevents left and right states from mixing and thereby equilibrating on even the longest time scales probed.  A high barrier acts effectively as an \textit{internal constraint} \cite{callen85,Gavrilov16a}.  
	Figure~\ref{fig:trajectories} shows our erasure protocols and four two-dimensional histograms of particle trajectories.

	Our protocol first adjusts the size of the right well to make the potential symmetric and brings left and right states into equilibrium.  It then erases a symmetric potential \cite{Jun14}:  lower down the barrier, allow the two states to mix, tilt the potential in the chosen direction, and raise back the barrier.  Lastly, the size of the right well is adjusted to its original.  
	For this cyclic erasure protocol, the particle initially has equal probability to be in either well and ends with probability $p_f=1$ to be in the chosen well.  That is, the one-bit memory is erased with probability $1$.  From the recorded positions $\bar{x}_n$ and trapping potential $U(x, t_n)$, we estimate work values for a fixed cycle time $\tau$.

\begin{figure}[t]
		\begin{center}
			\includegraphics[width=8.6cm]{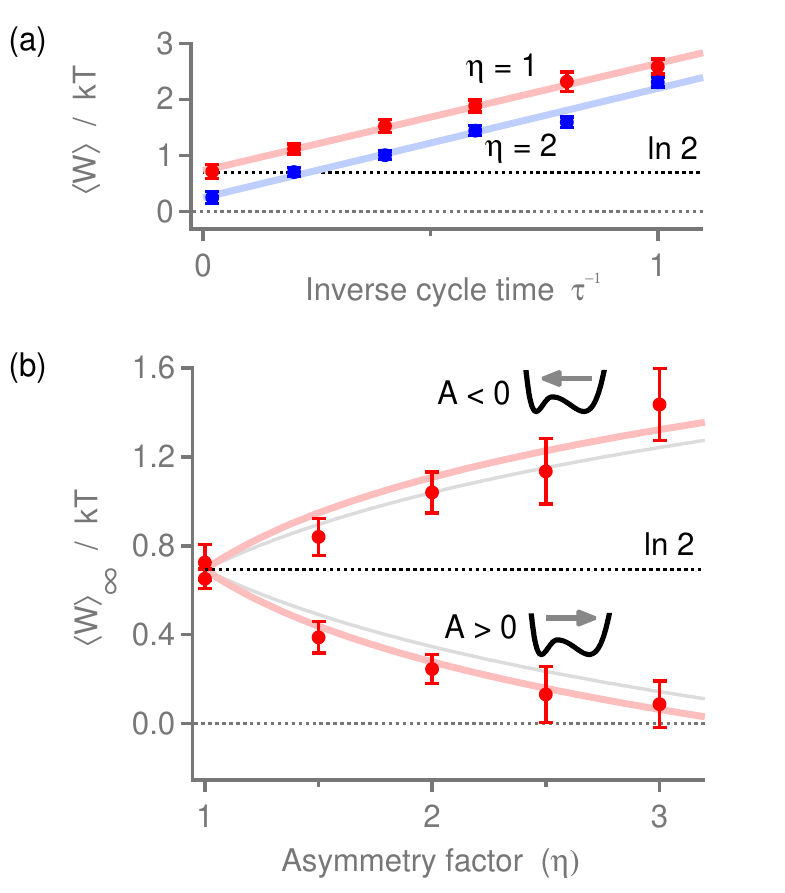}
		\end{center}
		\caption { \label{fig:WorkVsEta} 
			(color online) Mean work to erase bit depends on the symmetry of the potential.  (a) Work to erase the symmetric ($\eta=1$) and asymmetric ($\eta = 2$) bit. Solid lines show fit to the  asymptotic $\tau^{-1}$ correction.  
			The asymptotic work $\left<W\right>_\infty$ is the $y$-intercept.  (b) Asymptotic work as a function of the asymmetry factor $\eta$, when information is erased by resetting to the larger ($A>0$) or smaller ($A<0$) wells.  Thin gray line shows the asymptotic work prediction in Eq.~\ref{eq:WorkPrediction1}, while the thicker red line includes corrections due to the finite time step ($\Delta t = 0.005$ s).  Error bars are derived from least-squares fits in (a).}
	\end{figure}

	\noindent\textit{Result.} Figure~\ref{fig:WorkVsEta}a shows the mean erasure work as a function of inverse cycle time $\tau^{-1}$ for two different cases: erasure in a symmetric potential and erasure to the larger well ($\eta=2$).  From the finite cycle times, we estimate work in an arbitrarily slow limit, as the y-axis intercept.  The solid lines are fits to the expected asymptotic form $\left <W\right >_{\tau}\sim\left<W\right>_{\infty}+a\tau^{-1}$ \cite{sekimoto97a,schmiedl08}, where $a$ is a protocol-dependent constant.  
	We find that the asymptotic work to erase a symmetric memory is $\left<W\right>_{\infty}/kT=0.72 \pm 0.08$, which is comparable with the Landauer limit, $\ln 2\approx 0.69$.  However, the asymptotic work to erase an asymmetric memory to the larger well is $0.25~\pm~0.07$, significantly below $\ln 2$.  The asymptotic work, $\left<W\right>_{\infty}$ is estimated for several different values of $\eta$, ranging from 1 to 3 and plotted in Fig.~\ref{fig:WorkVsEta}b. When the memory is erased by resetting to the larger well ($A>0$), the measured work is below $kT\ln2$; by contrast, it is above the limit when erasing to the smaller well ($A<0$).
	
	\noindent\textit{Analysis.} As predicted \cite{shizume95,fahn96,barkeshli05,Sagawa09,turgut09,KlagesBook13, sagawa14,boyd16}, we observe that the asymptotic mean work deviates from $\ln 2$.  We then follow the work of Sagawa and Ueda \cite{Sagawa09,sagawa14} to calculate the asymptotic work values as a function of $\eta$.  The protocol in Fig.~\ref{fig:trajectories} can be decomposed into three basic operations:  compression of one well, erasure of a symmetric bit, and expansion of the compressed well.  Initially, the number of possible microstates is compressed by a factor of $\eta$, but only in half the cases; in the other half, the particles are in the unchanged well.  If the large well is occupied, the work to compress it quasistatically by a factor of $\eta$ is $\left<W_1\right>/kT = \tfrac{1}{2}\ln\eta^2 = \ln\eta$.  The work to erase a symmetric one-bit memory is exactly $\left<W_2\right>/kT = \ln 2$ \cite{landauer61}, for our erasure protocol.  Finally, expanding the potential quasistatically to its original shape extracts a work $\left< W_3 \right> /kT = \tfrac{1}{2} \ln \eta^{-2} = -\ln\eta$.  
	
	To compute the total work to erase an asymmetric memory, we must weight work components by their particle-occupation probability.  Since compression occurs in half the cases, we weight $\left<W_1\right>$ by a factor $\tfrac{1}{2}$.  Then, since symmetric erasure occurs in all cases, $\left<W_2\right>$ is weighted by 1.  Finally, in the expansion phase, the well is either always or never occupied, depending on which well we erase to.  The average work to erase a one-bit memory in a continuous, smooth, asymmetric double-well potential is then bounded from below by
	\begin{align}
		\frac{\left< W \right>}{kT} 
		&= \frac{1}{2}\frac{\left<W_1\right>}{kT} + \frac{\left<W_2\right>}{kT} + \left(\frac{1}{2}\pm\frac{1}{2} \right)\frac{\left<W_3\right>}{kT}\nonumber \\
		&= \ln2 \pm \tfrac{1}{2}\ln\eta \,,
		\label{eq:WorkPrediction1}
	\end{align}
	which is plotted in Fig.~\ref{fig:WorkVsEta}b (thin gray lines).
	
	Our use of a virtual potential leads to small corrections in the prediction of average work 
	(see Supplemental Material \footnote{
		See Supplemental Material [url], which includes Refs. \cite{Gavrilov16a,weigel14,gavrilov15,berglund08,gavrilov13,sekimoto97,sekimoto10,happel83,jun12,Jun14,gavrilov14,kawai07,parrondo09,roldan14,chiuchiu15,zulkowski14,dillenschneider09,callen85,KlagesBook13}
	}).   
	The corrected curves are shown as solid red lines in Fig.~\ref{fig:WorkVsEta}b.
	
	Equation~\ref{eq:WorkPrediction1} may also be interpreted as a generalized Landauer principle \cite{KlagesBook13}, where the work to erase a one-bit memory,
	\begin{equation}
		W_{\rm erase} \geq kT H - \Delta F \,,
		\label{eq:genLandauer}
	\end{equation}
	equals or exceeds the change in Shannon information, $H = \ln 2$, plus the average change in free energy between the final and each initial state.  Because of the high barrier, the initial and final states are locally in equilibrium, with an average free energy difference of $\Delta F /kT = \mp \frac{1}{2} \ln \eta$.  Equation~\ref{eq:genLandauer} becomes an equality for quasistatic transformations and from it, we also derive Eq.~\ref{eq:WorkPrediction1} in \cite{Note1}.

\begin{figure}[tb]
		\begin{center}
			\includegraphics[width=8.6cm]{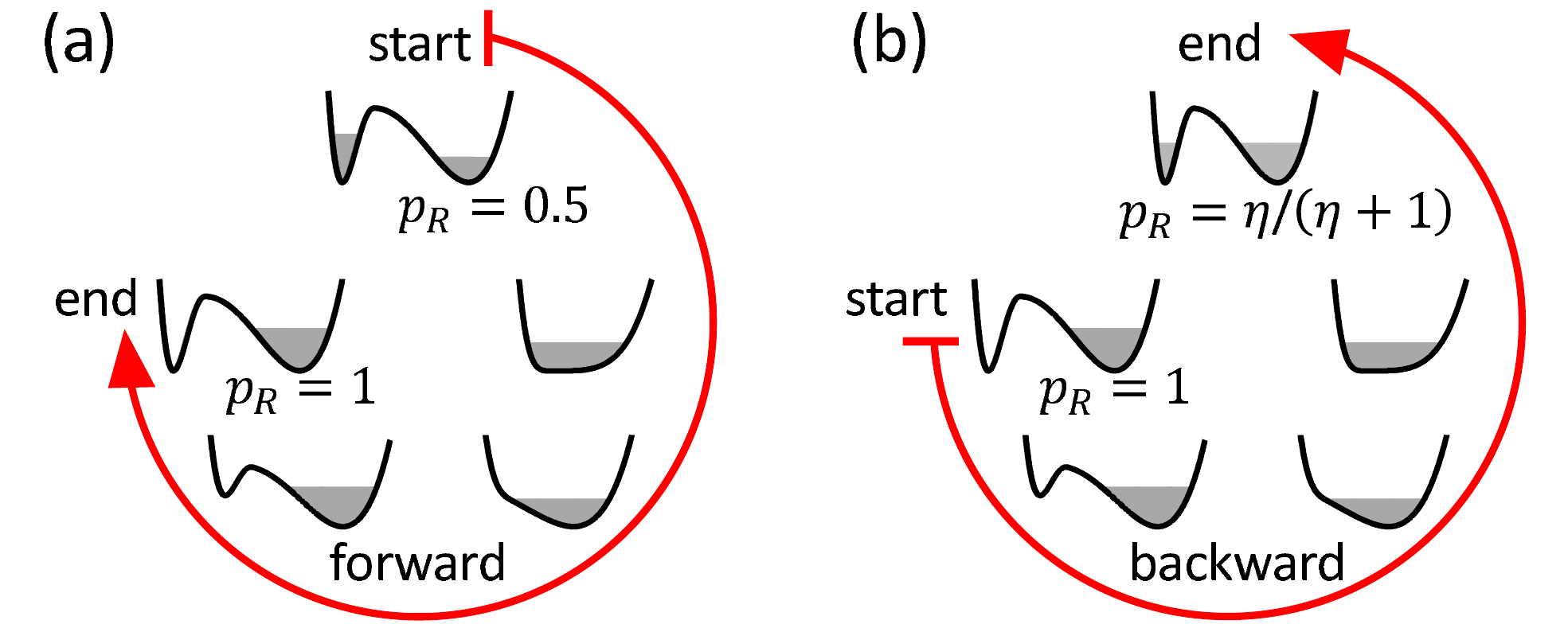}
		\end{center}
		\caption { \label{fig:IrrProtocol} 
			(color online) Thermodynamically irreversible erasure protocol.  (a) Forward protocol (clockwise):  lower the barrier, tilt, raise the barrier, and untilt.  Particles are distributed evenly ($p=0.5$) in either well but end up in the right well with probability $p=1$, as in the reversible protocol in Fig.~\ref{fig:trajectories}.  (b)  Backward protocol (counterclockwise):  a particle that starts with probability $p=1$ in the right well and executes the protocol in reverse ends up in the right well with probability $p=\eta/(\eta +1)$, which differs from 0.5 when $\eta > 1$.}
	\end{figure}

\begin{figure}[tb]
		\begin{center}
			\includegraphics[width=8.6cm]{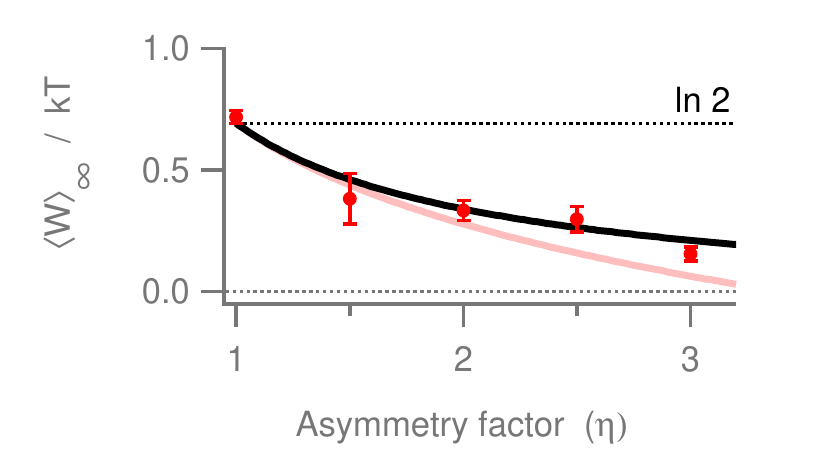}
		\end{center}
		\caption { \label{fig:WorkIrreversible} 
			(color online) The thermodynamically irreversible forward protocol requires more work (markers) than the prediction based on a reversible process (Eq.~\ref{eq:WorkPrediction1}, thin light line).  A calculation accounting for irreversibility (heavy darker line) is consistent with the data \cite{Note1}.}
	\end{figure}

	\noindent\textit{Thermodynamic and logical irreversibility.} Although Eq.~\ref{eq:WorkPrediction1} would seem to be a universal result for the asymptotic, average work to erase an asymmetric potential, one must be cautious.  Indeed, as shown in Fig.~\ref{fig:IrrProtocol}a, we first considered what might seem to be an equivalent protocol; however, we found that the observed asymptotic work (Fig.~\ref{fig:WorkIrreversible}) differs markedly from the prediction based on Eq.~\ref{eq:WorkPrediction1} (markers).  In this case, the corrections due to finite feedback times are \textit{not} responsible.

	To understand the difference between the two protocols, we first recall that logical and thermodynamic reversibility denote two distinct concepts \cite{maroney05,sagawa14,Lopez16}.   Here, both protocols are \textit{logically irreversible}:  starting from an even probability to be in either well, the particle will end up, with certainty, in the well chosen for erasure.  The protocols are not logically reversible because the output state does not uniquely define the input state  \cite{landauer61}.
	
	By contrast, \textit{thermodynamic reversibility} asks whether the sequence of thermodynamic state (occupation  probabilities) is the same in the forward and backwards protocols \cite{callen85}.  It is easy to see from Fig.~\ref{fig:trajectories} that the original protocol is reversible in this sense.  In the original protocol, erasing one bit of entropy originally associated with the logical degrees of freedom---localizing the particle to a single well---transfers an equivalent amount of heat in the bath in a way that does not increase the entropy of the Universe.  Reversing the process merely removes the same amount of heat from the bath and moves its equivalent to the logical degrees of freedom---\textit{creating} information in the ``information-bearing degrees of freedom" \cite{boyd16}, a point often forgotten \cite{maroney09}.  
	
	Although the second protocol is logically equivalent to the first---one bit of system entropy is fully erased---it is different as regards the thermodynamic reversibility.  Figure~\ref{fig:IrrProtocol} illustrates the point by showing that the occupation probabilities for the forward sequence of states (a) differs from that of the time-reversed sequence of states (b).  Thus, the second protocol is not reversible and not uniformly quasistatic, even though it is performed much more slowly than the relevant time scales of the problem.  (As we mentioned above, the raised  barrier is high enough to act as an internal constraint that divides the phase space into disjoint regions.)  
	
\begin{figure}[tb]
		\begin{center}
			\includegraphics[width=8.6cm]{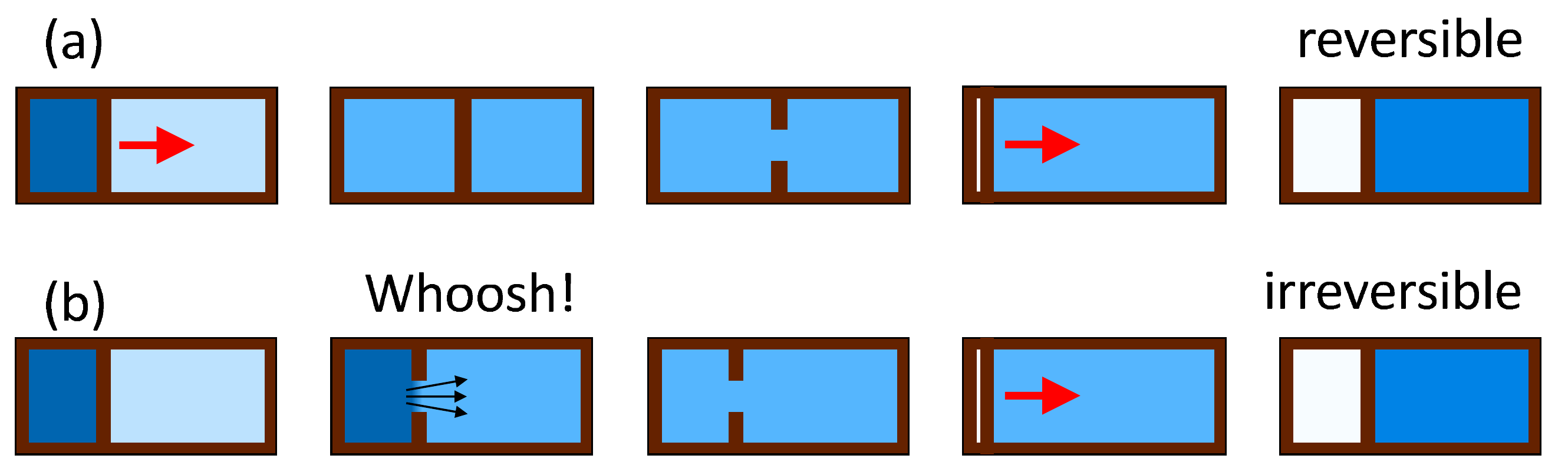}
		\end{center}
		\caption { \label{fig:GasAnalogyErasure} 
			(color online) Ideal gas in closed container with two chambers is analogous to an asymmetric memory.  Valve on divider can open to mix gases. (a) Reversible protocol equalizes pressures, mixes gases, and compresses to one vessel.  Starting from the final state and slowly transforming back will reproduce the initial state.  (b) Irreversible protocol mixes the gases by free expansion---with a ``whoosh''---and then compresses.  Starting from the final state and slowly transforming back will not reproduce the initial state.
		}
	\end{figure}

	\noindent\textit{Macroscopic reversibility.} This property of irreversibility in a slow transformation is analogous to that of a free expansion of an ideal gas \cite{callen85,Gavrilov16a,Note1}.  Figure \ref{fig:GasAnalogyErasure} shows the point using a schematic of an ideal-gas system analogous to the asymmetric-erasure experiment:  Each vessel contains the same number of ideal-gas molecules. One vessel is $\eta$ times larger resulting in $\eta$ times lower pressure (light blue).
	The asymmetric-erasure experiment is analogous to the compression of an ideal gas into one vessel.  Such a compression can be achieved in thermodynamically reversible (a) and irreversible (b) ways.  The reversible protocol first makes the system ergodic by moving the separator to equalize pressures; then it mixes gases and compresses to the larger volume.  The irreversible protocol mixes two gases at different pressures directly, which leads to the irreversible process of free expansion.  In the end, the state of the ideal gas is the same for both protocols, but more work required is for the irreversible case. 
	
	\noindent\textit{Mesoscopic reversibility.} The notion of an arbitrarily slow yet irreversible protocol is subtle at the level of single-molecule studies.  Just as no small-scale experiment can demonstrate a thermodynamic law in a single cycle, no single experiment on a mesoscopic system can establish reversibility.  But compiling statistics over many forward and backward realizations can test reversibility \cite{roldan15}.  Indeed, in re-examining the work of Dillenschneider and Lutz that first advocated testing Landauer's principle in a small system  \cite{dillenschneider09}, we noticed that their protocol is in fact irreversible under the conditions they studied:  The tilt is started before lowering the barrier, leading to an intrinsically irreversible transformation.  Delaying the start of the tilt corrects the problem.  See Supplemental material \cite{Note1}.
	
	\noindent \textit{Conclusion}.  By studying erasure in an asymmetric double-well virtual potential, we have confirmed the prediction that information erasure can be accomplished in a mesoscopic system using a mean work that is less than $kT$ $\ln 2$ per bit.  Whether this ``violates" or ``generalizes" Landauer's principle is perhaps more a matter of semantics than physics.  Physically, the reduced work arises when the starting state is not in equilibrium, and other degrees of freedom do work that compensates the work required to erase.  More simply, erasing from a small well to a large well  transfers a particle from a small box to a larger one, but never the reverse.  The transfer is a single-particle, mesoscopic version of gas expansion against a piston.  The net reduction of work can be understood from the succession of quasistatic states, as long as one accounts for all the degrees of freedom---the information-bearing degrees of freedom (macrostates), the differing number of microstates associated with each macrostate, and the microstates associated with the heat bath.  More subtly, by considering two slightly different protocols that erase information identically but are  different thermodynamically, we can clarify the relationship between logical and thermodynamic irreversibility, an issue that can still confuse, many years after the contributions of Maxwell, Szilard, Landauer, and Bennett.

	\begin{acknowledgments}
		We thank Jordan Horowitz for helpful discussions.  This work was supported by NSERC (Canada).
	\end{acknowledgments}


\begin{thebibliography}{61}%
	\makeatletter
	\providecommand \@ifxundefined [1]{%
		\@ifx{#1\undefined}
	}%
	\providecommand \@ifnum [1]{%
		\ifnum #1\expandafter \@firstoftwo
		\else \expandafter \@secondoftwo
		\fi
	}%
	\providecommand \@ifx [1]{%
		\ifx #1\expandafter \@firstoftwo
		\else \expandafter \@secondoftwo
		\fi
	}%
	\providecommand \natexlab [1]{#1}%
	\providecommand \enquote  [1]{``#1''}%
	\providecommand \bibnamefont  [1]{#1}%
	\providecommand \bibfnamefont [1]{#1}%
	\providecommand \citenamefont [1]{#1}%
	\providecommand \href@noop [0]{\@secondoftwo}%
	\providecommand \href [0]{\begingroup \@sanitize@url \@href}%
	\providecommand \@href[1]{\@@startlink{#1}\@@href}%
	\providecommand \@@href[1]{\endgroup#1\@@endlink}%
	\providecommand \@sanitize@url [0]{\catcode `\\12\catcode `\$12\catcode
		`\&12\catcode `\#12\catcode `\^12\catcode `\_12\catcode `\%12\relax}%
	\providecommand \@@startlink[1]{}%
	\providecommand \@@endlink[0]{}%
	\providecommand \url  [0]{\begingroup\@sanitize@url \@url }%
	\providecommand \@url [1]{\endgroup\@href {#1}{\urlprefix }}%
	\providecommand \urlprefix  [0]{URL }%
	\providecommand \Eprint [0]{\href }%
	\providecommand \doibase [0]{http://dx.doi.org/}%
	\providecommand \selectlanguage [0]{\@gobble}%
	\providecommand \bibinfo  [0]{\@secondoftwo}%
	\providecommand \bibfield  [0]{\@secondoftwo}%
	\providecommand \translation [1]{[#1]}%
	\providecommand \BibitemOpen [0]{}%
	\providecommand \bibitemStop [0]{}%
	\providecommand \bibitemNoStop [0]{.\EOS\space}%
	\providecommand \EOS [0]{\spacefactor3000\relax}%
	\providecommand \BibitemShut  [1]{\csname bibitem#1\endcsname}%
	\let\auto@bib@innerbib\@empty
	\bibitem [{\citenamefont {Landauer}(1961)}]{landauer61}%
	\BibitemOpen
	\bibfield  {author} {\bibinfo {author} {\bibfnamefont {R.}~\bibnamefont
			{Landauer}},\ }\bibfield  {title} {\enquote {\bibinfo {title}
			{Irreversibility and heat generation in the computing process},}\ }\href@noop
	{} {\bibfield  {journal} {\bibinfo  {journal} {IBM J. Res. Develop.}\
		}\textbf {\bibinfo {volume} {5}},\ \bibinfo {pages} {183--191} (\bibinfo
		{year} {1961})}\BibitemShut {NoStop}%
	\bibitem [{\citenamefont {Sagawa}(2014)}]{sagawa14}%
	\BibitemOpen
	\bibfield  {author} {\bibinfo {author} {\bibfnamefont {T.}~\bibnamefont
			{Sagawa}},\ }\bibfield  {title} {\enquote {\bibinfo {title} {Thermodynamic
				and logical reversibilities revisited},}\ }\href@noop {} {\bibfield
		{journal} {\bibinfo  {journal} {J. Stat. Mech.}\ ,\ \bibinfo {pages}
			{P03025}} (\bibinfo {year} {2014})}\BibitemShut {NoStop}%
	\bibitem [{\citenamefont {Maxwell}(1871)}]{maxwell1871}%
	\BibitemOpen
	\bibfield  {author} {\bibinfo {author} {\bibfnamefont {J.~C.}\ \bibnamefont
			{Maxwell}},\ }\href@noop {} {\emph {\bibinfo {title} {Theory of Heat}}}\
	(\bibinfo  {publisher} {Longmans, Green, and Co.},\ \bibinfo {year}
	{1871})\BibitemShut {NoStop}%
	\bibitem [{\citenamefont {Szilard}(1929)}]{szilard29}%
	\BibitemOpen
	\bibfield  {author} {\bibinfo {author} {\bibfnamefont {L.}~\bibnamefont
			{Szilard}},\ }\bibfield  {title} {\enquote {\bibinfo {title} {On the decrease
				of entropy in a thermodynamic system by the intervention of intelligent
				beings},}\ }\href@noop {} {\bibfield  {journal} {\bibinfo  {journal} {Z.
				Physik}\ }\textbf {\bibinfo {volume} {53}},\ \bibinfo {pages} {840--856}
		(\bibinfo {year} {1929})}\BibitemShut {NoStop}%
	\bibitem [{\citenamefont {Bennett}(1982)}]{bennett82}%
	\BibitemOpen
	\bibfield  {author} {\bibinfo {author} {\bibfnamefont {C.~H.}\ \bibnamefont
			{Bennett}},\ }\bibfield  {title} {\enquote {\bibinfo {title} {The
				thermodynamics of computation: a review},}\ }\href@noop {} {\bibfield
		{journal} {\bibinfo  {journal} {Int. J. Theor. Phys.}\ }\textbf {\bibinfo
			{volume} {21}},\ \bibinfo {pages} {905--940} (\bibinfo {year}
		{1982})}\BibitemShut {NoStop}%
	\bibitem [{\citenamefont {Leff}\ and\ \citenamefont {Rex}(2003)}]{leff03}%
	\BibitemOpen
	\bibfield  {author} {\bibinfo {author} {\bibfnamefont {H.~S.}\ \bibnamefont
			{Leff}}\ and\ \bibinfo {author} {\bibfnamefont {A.~F.}\ \bibnamefont {Rex}},\
	}\href@noop {} {\emph {\bibinfo {title} {Maxwell's Demon 2: Entropy,
			Classical and Quantum Information, Computing}}}\ (\bibinfo  {publisher}
{IOP},\ \bibinfo {year} {2003})\BibitemShut {NoStop}%
\bibitem [{\citenamefont {B{\'e}rut}\ \emph {et~al.}(2012)\citenamefont
	{B{\'e}rut}, \citenamefont {Arakelyan}, \citenamefont {Petrosyan},
	\citenamefont {Ciliberto}, \citenamefont {Dillenschneider},\ and\
	\citenamefont {Lutz}}]{berut12}%
\BibitemOpen
\bibfield  {author} {\bibinfo {author} {\bibfnamefont {A.}~\bibnamefont
		{B{\'e}rut}}, \bibinfo {author} {\bibfnamefont {A.}~\bibnamefont
		{Arakelyan}}, \bibinfo {author} {\bibfnamefont {A.}~\bibnamefont
		{Petrosyan}}, \bibinfo {author} {\bibfnamefont {S.}~\bibnamefont
		{Ciliberto}}, \bibinfo {author} {\bibfnamefont {R.}~\bibnamefont
		{Dillenschneider}}, \ and\ \bibinfo {author} {\bibfnamefont {E.}~\bibnamefont
		{Lutz}},\ }\bibfield  {title} {\enquote {\bibinfo {title} {Experimental
			verification of {L}andauer's principle linking information and
			thermodynamics},}\ }\href@noop {} {\bibfield  {journal} {\bibinfo  {journal}
		{Nature}\ }\textbf {\bibinfo {volume} {483}},\ \bibinfo {pages} {187--190}
	(\bibinfo {year} {2012})}\BibitemShut {NoStop}%
\bibitem [{\citenamefont {Jun}\ \emph {et~al.}(2014)\citenamefont {Jun},
	\citenamefont {Gavrilov},\ and\ \citenamefont {Bechhoefer}}]{Jun14}%
\BibitemOpen
\bibfield  {author} {\bibinfo {author} {\bibfnamefont {Y.}~\bibnamefont
		{Jun}}, \bibinfo {author} {\bibfnamefont {M.}~\bibnamefont {Gavrilov}}, \
	and\ \bibinfo {author} {\bibfnamefont {J.}~\bibnamefont {Bechhoefer}},\
}\bibfield  {title} {\enquote {\bibinfo {title} {High-precision test of
		{L}andauer's principle in a feedback trap},}\ }\href@noop {} {\bibfield
{journal} {\bibinfo  {journal} {Phys. Rev. Lett.}\ }\textbf {\bibinfo
	{volume} {113}},\ \bibinfo {pages} {190601} (\bibinfo {year}
{2014})}\BibitemShut {NoStop}%
\bibitem [{\citenamefont {Hong}\ \emph {et~al.}(2016)\citenamefont {Hong},
	\citenamefont {Lambson}, \citenamefont {Dhuey},\ and\ \citenamefont
	{Bokor}}]{Hong16}%
\BibitemOpen
\bibfield  {author} {\bibinfo {author} {\bibfnamefont {J.}~\bibnamefont
		{Hong}}, \bibinfo {author} {\bibfnamefont {B.}~\bibnamefont {Lambson}},
	\bibinfo {author} {\bibfnamefont {S.}~\bibnamefont {Dhuey}}, \ and\ \bibinfo
	{author} {\bibfnamefont {J.}~\bibnamefont {Bokor}},\ }\bibfield  {title}
{\enquote {\bibinfo {title} {Experimental test of {L}andauer's principle in
			single-bit operations on nanomagnetic memory bits},}\ }\href@noop {}
{\bibfield  {journal} {\bibinfo  {journal} {Sci. Adv.}\ }\textbf {\bibinfo
		{volume} {2}},\ \bibinfo {pages} {e1501492} (\bibinfo {year}
	{2016})}\BibitemShut {NoStop}%
\bibitem [{\citenamefont {Martini}\ \emph {et~al.}(2016)\citenamefont
	{Martini}, \citenamefont {Pancaldi}, \citenamefont {Madami}, \citenamefont
	{Vavassori}, \citenamefont {Gubbiotti}, \citenamefont {Tacchi}, \citenamefont
	{Hartmann}, \citenamefont {Emmerling}, \citenamefont {H{\"o}fling},
	\citenamefont {Worschech},\ and\ \citenamefont {Carlotti}}]{Martini16}%
\BibitemOpen
\bibfield  {author} {\bibinfo {author} {\bibfnamefont {L.}~\bibnamefont
		{Martini}}, \bibinfo {author} {\bibfnamefont {M.}~\bibnamefont {Pancaldi}},
	\bibinfo {author} {\bibfnamefont {M.}~\bibnamefont {Madami}}, \bibinfo
	{author} {\bibfnamefont {P.}~\bibnamefont {Vavassori}}, \bibinfo {author}
	{\bibfnamefont {G.}~\bibnamefont {Gubbiotti}}, \bibinfo {author}
	{\bibfnamefont {S.}~\bibnamefont {Tacchi}}, \bibinfo {author} {\bibfnamefont
		{F.}~\bibnamefont {Hartmann}}, \bibinfo {author} {\bibfnamefont
		{M.}~\bibnamefont {Emmerling}}, \bibinfo {author} {\bibfnamefont
		{S.}~\bibnamefont {H{\"o}fling}}, \bibinfo {author} {\bibfnamefont
		{L.}~\bibnamefont {Worschech}}, \ and\ \bibinfo {author} {\bibfnamefont
		{G.}~\bibnamefont {Carlotti}},\ }\bibfield  {title} {\enquote {\bibinfo
		{title} {Experimental and theoretical analysis of {L}andauer erasure in
			nano-magnetic switches of different sizes},}\ }\href@noop {} {\bibfield
	{journal} {\bibinfo  {journal} {Nano Energy}\ }\textbf {\bibinfo {volume}
		{19}},\ \bibinfo {pages} {108 -- 116} (\bibinfo {year} {2016})}\BibitemShut
{NoStop}%
\bibitem [{\citenamefont {Peterson}\ \emph {et~al.}(2016)\citenamefont
	{Peterson}, \citenamefont {Sarthour}, \citenamefont {Souza}, \citenamefont
	{Oliveira}, \citenamefont {Goold}, \citenamefont {Modi}, \citenamefont
	{Soares-Pinto},\ and\ \citenamefont {C{\'e}leri}}]{Peterson16}%
\BibitemOpen
\bibfield  {author} {\bibinfo {author} {\bibfnamefont {J.~P.~S.}\
		\bibnamefont {Peterson}}, \bibinfo {author} {\bibfnamefont {R.~S.}\
		\bibnamefont {Sarthour}}, \bibinfo {author} {\bibfnamefont {A.~M.}\
		\bibnamefont {Souza}}, \bibinfo {author} {\bibfnamefont {I.~S.}\ \bibnamefont
		{Oliveira}}, \bibinfo {author} {\bibfnamefont {J.}~\bibnamefont {Goold}},
	\bibinfo {author} {\bibfnamefont {K.}~\bibnamefont {Modi}}, \bibinfo {author}
	{\bibfnamefont {D.~O.}\ \bibnamefont {Soares-Pinto}}, \ and\ \bibinfo
	{author} {\bibfnamefont {L.~C.}\ \bibnamefont {C{\'e}leri}},\ }\bibfield
{title} {\enquote {\bibinfo {title} {Experimental demonstration of
			information to energy conversion in a quantum system at the {L}andauer
			limit},}\ }\href@noop {} {\bibfield  {journal} {\bibinfo  {journal} {Proc. R.
			Soc. A}\ }\textbf {\bibinfo {volume} {472}},\ \bibinfo {pages} {2015.0813}
	(\bibinfo {year} {2016})}\BibitemShut {NoStop}%
\bibitem [{\citenamefont {Sekimoto}(2010)}]{sekimoto10}%
\BibitemOpen
\bibfield  {author} {\bibinfo {author} {\bibfnamefont {K.}~\bibnamefont
		{Sekimoto}},\ }\href@noop {} {\emph {\bibinfo {title} {Stochastic
			Energetics}}}\ (\bibinfo  {publisher} {Springer},\ \bibinfo {year}
{2010})\BibitemShut {NoStop}%
\bibitem [{\citenamefont {Seifert}(2012)}]{seifert12}%
\BibitemOpen
\bibfield  {author} {\bibinfo {author} {\bibfnamefont {U.}~\bibnamefont
		{Seifert}},\ }\bibfield  {title} {\enquote {\bibinfo {title} {Stochastic
			thermodynamics, fluctuation theorems and molecular machines},}\ }\href@noop
{} {\bibfield  {journal} {\bibinfo  {journal} {Rep. Prog. Phys.}\ }\textbf
	{\bibinfo {volume} {75}},\ \bibinfo {pages} {1--58} (\bibinfo {year}
	{2012})}\BibitemShut {NoStop}%
\bibitem [{\citenamefont {Parrondo}\ \emph {et~al.}(2015)\citenamefont
	{Parrondo}, \citenamefont {Horowitz},\ and\ \citenamefont
	{Sagawa}}]{parrondo15}%
\BibitemOpen
\bibfield  {author} {\bibinfo {author} {\bibfnamefont {Juan M.~R.}\
		\bibnamefont {Parrondo}}, \bibinfo {author} {\bibfnamefont {Jordan~M.}\
		\bibnamefont {Horowitz}}, \ and\ \bibinfo {author} {\bibfnamefont {Takahiro}\
		\bibnamefont {Sagawa}},\ }\bibfield  {title} {\enquote {\bibinfo {title}
		{Thermodynamics of information},}\ }\href@noop {} {\bibfield  {journal}
	{\bibinfo  {journal} {Nature Physics}\ } (\bibinfo {year}
	{2015})}\BibitemShut {NoStop}%
\bibitem [{\citenamefont {Toyabe}\ \emph {et~al.}(2010)\citenamefont {Toyabe},
	\citenamefont {Sagawa}, \citenamefont {Ueda}, \citenamefont {Muneyuki},\ and\
	\citenamefont {Sano}}]{toyabe10}%
\BibitemOpen
\bibfield  {author} {\bibinfo {author} {\bibfnamefont {S.}~\bibnamefont
		{Toyabe}}, \bibinfo {author} {\bibfnamefont {T.}~\bibnamefont {Sagawa}},
	\bibinfo {author} {\bibfnamefont {M.}~\bibnamefont {Ueda}}, \bibinfo {author}
	{\bibfnamefont {E.}~\bibnamefont {Muneyuki}}, \ and\ \bibinfo {author}
	{\bibfnamefont {M.}~\bibnamefont {Sano}},\ }\bibfield  {title} {\enquote
	{\bibinfo {title} {Experimental demonstration of information-to-energy
			conversion and validation of the generalized {J}arzynski equality},}\
}\href@noop {} {\bibfield  {journal} {\bibinfo  {journal} {Nature Phys.}\
}\textbf {\bibinfo {volume} {6}},\ \bibinfo {pages} {988} (\bibinfo {year}
{2010})}\BibitemShut {NoStop}%
\bibitem [{\citenamefont {Koski}\ \emph {et~al.}(2014)\citenamefont {Koski},
	\citenamefont {Maisi}, \citenamefont {Pekola},\ and\ \citenamefont
	{Averin}}]{koski14a}%
\BibitemOpen
\bibfield  {author} {\bibinfo {author} {\bibfnamefont {J.~V.}\ \bibnamefont
		{Koski}}, \bibinfo {author} {\bibfnamefont {V.~F.}\ \bibnamefont {Maisi}},
	\bibinfo {author} {\bibfnamefont {J.~P.}\ \bibnamefont {Pekola}}, \ and\
	\bibinfo {author} {\bibfnamefont {D.~M.}\ \bibnamefont {Averin}},\ }\bibfield
{title} {\enquote {\bibinfo {title} {Experimental realization of a {S}zilard
			engine with a single electron},}\ }\href@noop {} {\bibfield  {journal}
	{\bibinfo  {journal} {Proc. Natl. Acad. Sci. USA}\ }\textbf {\bibinfo
		{volume} {111}},\ \bibinfo {pages} {13786--13789} (\bibinfo {year}
	{2014})}\BibitemShut {NoStop}%
\bibitem [{\citenamefont {Koski}\ \emph {et~al.}(2015)\citenamefont {Koski},
	\citenamefont {Kutvonen}, \citenamefont {Khaymovich}, \citenamefont
	{Ala-Nissila},\ and\ \citenamefont {Pekola}}]{koski15}%
\BibitemOpen
\bibfield  {author} {\bibinfo {author} {\bibfnamefont {J.~V.}\ \bibnamefont
		{Koski}}, \bibinfo {author} {\bibfnamefont {A.}~\bibnamefont {Kutvonen}},
	\bibinfo {author} {\bibfnamefont {I.~M.}\ \bibnamefont {Khaymovich}},
	\bibinfo {author} {\bibfnamefont {T.}~\bibnamefont {Ala-Nissila}}, \ and\
	\bibinfo {author} {\bibfnamefont {J.~P.}\ \bibnamefont {Pekola}},\ }\bibfield
{title} {\enquote {\bibinfo {title} {On-chip {M}axwell's demon as an
			information-powered refrigerator},}\ }\href@noop {} {\bibfield  {journal}
	{\bibinfo  {journal} {Phys. Rev. Lett.}\ }\textbf {\bibinfo {volume} {115}},\
	\bibinfo {pages} {260602} (\bibinfo {year} {2015})}\BibitemShut {NoStop}%
\bibitem [{\citenamefont {Camati}\ \emph {et~al.}(2016)\citenamefont {Camati},
	\citenamefont {Peterson}, \citenamefont {Batalh{\~a}o}, \citenamefont
	{Micadei}, \citenamefont {Souza}, \citenamefont {Sarthour}, \citenamefont
	{Oliveira},\ and\ \citenamefont {Serra}}]{camati16}%
\BibitemOpen
\bibfield  {author} {\bibinfo {author} {\bibfnamefont {P.~A.}\ \bibnamefont
		{Camati}}, \bibinfo {author} {\bibfnamefont {J.~P.~S.}\ \bibnamefont
		{Peterson}}, \bibinfo {author} {\bibfnamefont {T.~B.}\ \bibnamefont
		{Batalh{\~a}o}}, \bibinfo {author} {\bibfnamefont {K.}~\bibnamefont
		{Micadei}}, \bibinfo {author} {\bibfnamefont {A.~M.}\ \bibnamefont {Souza}},
	\bibinfo {author} {\bibfnamefont {R.~S.}\ \bibnamefont {Sarthour}}, \bibinfo
	{author} {\bibfnamefont {I.~S.}\ \bibnamefont {Oliveira}}, \ and\ \bibinfo
	{author} {\bibfnamefont {R.~M.}\ \bibnamefont {Serra}},\ }\bibfield  {title}
{\enquote {\bibinfo {title} {Experimental rectification of entropy production
			by a {M}axwell's {D}emon in a quantum system},}\ }\href@noop {} {\ ,\
	\bibinfo {pages} {arxiv1605.08821} (\bibinfo {year} {2016})}\BibitemShut
{NoStop}%
\bibitem [{\citenamefont {Shizume}(1995)}]{shizume95}%
\BibitemOpen
\bibfield  {author} {\bibinfo {author} {\bibfnamefont {Kousuke}\ \bibnamefont
		{Shizume}},\ }\bibfield  {title} {\enquote {\bibinfo {title} {Heat generation
			required by information erasure},}\ }\href@noop {} {\bibfield  {journal}
	{\bibinfo  {journal} {Phys. Rev. E}\ }\textbf {\bibinfo {volume} {52}},\
	\bibinfo {pages} {3495--3499} (\bibinfo {year} {1995})}\BibitemShut {NoStop}%
\bibitem [{\citenamefont {Fahn}(1996)}]{fahn96}%
\BibitemOpen
\bibfield  {author} {\bibinfo {author} {\bibfnamefont {Paul~N.}\ \bibnamefont
		{Fahn}},\ }\bibfield  {title} {\enquote {\bibinfo {title} {Maxwell's demon
			and the entropy cost of information},}\ }\href@noop {} {\bibfield  {journal}
	{\bibinfo  {journal} {Foundations of Physics}\ }\textbf {\bibinfo {volume}
		{26}},\ \bibinfo {pages} {71--93} (\bibinfo {year} {1996})}\BibitemShut
{NoStop}%
\bibitem [{\citenamefont {Barkeshli}(2005)}]{barkeshli05}%
\BibitemOpen
\bibfield  {author} {\bibinfo {author} {\bibfnamefont {M~M}\ \bibnamefont
		{Barkeshli}},\ }\bibfield  {title} {\enquote {\bibinfo {title}
		{Dissipationless information erasure and the breakdown of {L}andauer's
			principle},}\ }\href@noop {} {\ ,\ \bibinfo {pages} {arxiv:0504.323}
	(\bibinfo {year} {2005})}\BibitemShut {NoStop}%
\bibitem [{\citenamefont {Sagawa}\ and\ \citenamefont {Ueda}(2009)}]{Sagawa09}%
\BibitemOpen
\bibfield  {author} {\bibinfo {author} {\bibfnamefont {T.}~\bibnamefont
		{Sagawa}}\ and\ \bibinfo {author} {\bibfnamefont {M.}~\bibnamefont {Ueda}},\
}\bibfield  {title} {\enquote {\bibinfo {title} {Minimal energy cost for
		thermodynamic information processing: Measurement and information erasure},}\
}\href@noop {} {\bibfield  {journal} {\bibinfo  {journal} {Phys. Rev. Lett.}\
}\textbf {\bibinfo {volume} {102}},\ \bibinfo {pages} {250602} (\bibinfo
{year} {2009})}\BibitemShut {NoStop}%
\bibitem [{\citenamefont {Turgut}(2009)}]{turgut09}%
\BibitemOpen
\bibfield  {author} {\bibinfo {author} {\bibfnamefont {S.}~\bibnamefont
		{Turgut}},\ }\bibfield  {title} {\enquote {\bibinfo {title} {Relations
			between entropies produced in nondeterministic thermodynamic processes},}\
}\href@noop {} {\bibfield  {journal} {\bibinfo  {journal} {Phys. Rev. E}\
}\textbf {\bibinfo {volume} {79}},\ \bibinfo {pages} {041102} (\bibinfo
{year} {2009})}\BibitemShut {NoStop}%
\bibitem [{\citenamefont {Sagawa}\ and\ \citenamefont
	{Ueda}(2013)}]{KlagesBook13}%
\BibitemOpen
\bibfield  {author} {\bibinfo {author} {\bibfnamefont {Takahiro}\
		\bibnamefont {Sagawa}}\ and\ \bibinfo {author} {\bibfnamefont {Masahito}\
		\bibnamefont {Ueda}},\ }\href@noop {} {\emph {\bibinfo {title} {Information
			Thermodynamics: Maxwell's Demon in Nonequilibrium Dynamics}}},\ edited by\
\bibinfo {editor} {\bibfnamefont {Rainer}\ \bibnamefont {Klages}}, \bibinfo
{editor} {\bibfnamefont {Wolfram}\ \bibnamefont {Just}}, \ and\ \bibinfo
{editor} {\bibfnamefont {Christopher}\ \bibnamefont {Jarzynski}}\ (\bibinfo
{publisher} {Wiley-VCH, Weinheim, 2013},\ \bibinfo {year} {2013})\BibitemShut
{NoStop}%
\bibitem [{\citenamefont {Boyd}\ and\ \citenamefont
	{Crutchfield}(2016)}]{boyd16}%
\BibitemOpen
\bibfield  {author} {\bibinfo {author} {\bibfnamefont {Alexander~B.}\
		\bibnamefont {Boyd}}\ and\ \bibinfo {author} {\bibfnamefont {James~P.}\
		\bibnamefont {Crutchfield}},\ }\bibfield  {title} {\enquote {\bibinfo {title}
		{Maxwell demon dynamics: Deterministic chaos, the {S}zilard map, and the
			intelligence of thermodynamic systems},}\ }\href@noop {} {\bibfield
	{journal} {\bibinfo  {journal} {Phys. Rev. Lett.}\ }\textbf {\bibinfo
		{volume} {116}},\ \bibinfo {pages} {190601} (\bibinfo {year}
	{2016})}\BibitemShut {NoStop}%
\bibitem [{\citenamefont {Ouldridge}\ \emph {et~al.}(2016)\citenamefont
	{Ouldridge}, \citenamefont {Govern},\ and\ \citenamefont
	{Wolde}}]{ouldridge16}%
\BibitemOpen
\bibfield  {author} {\bibinfo {author} {\bibfnamefont {TE}~\bibnamefont
		{Ouldridge}}, \bibinfo {author} {\bibfnamefont {CC}~\bibnamefont {Govern}}, \
	and\ \bibinfo {author} {\bibfnamefont {PRT}\ \bibnamefont {Wolde}},\
}\bibfield  {title} {\enquote {\bibinfo {title} {The thermodynamics of
		computational copying in biochemical systems},}\ }\href@noop {} {\ ,\
\bibinfo {pages} {arxiv1503.00909v3} (\bibinfo {year} {2016})}\BibitemShut
{NoStop}%
\bibitem [{\citenamefont {Dillenschneider}\ and\ \citenamefont
	{Lutz}(2010)}]{Dill10}%
\BibitemOpen
\bibfield  {author} {\bibinfo {author} {\bibfnamefont {Raoul}\ \bibnamefont
		{Dillenschneider}}\ and\ \bibinfo {author} {\bibfnamefont {Eric}\
		\bibnamefont {Lutz}},\ }\bibfield  {title} {\enquote {\bibinfo {title}
		{Comment on `minimal energy cost for thermodynamic information processing:
			Measurement and information erasure'},}\ }\href@noop {} {\bibfield  {journal}
	{\bibinfo  {journal} {Phys. Rev. Lett.}\ }\textbf {\bibinfo {volume} {104}},\
	\bibinfo {pages} {198903} (\bibinfo {year} {2010})}\BibitemShut {NoStop}%
\bibitem [{\citenamefont {Sagawa}\ and\ \citenamefont {Ueda}(2010)}]{Sagawa10}%
\BibitemOpen
\bibfield  {author} {\bibinfo {author} {\bibfnamefont {Takahiro}\
		\bibnamefont {Sagawa}}\ and\ \bibinfo {author} {\bibfnamefont {Masahito}\
		\bibnamefont {Ueda}},\ }\bibfield  {title} {\enquote {\bibinfo {title}
		{Sagawa and {U}eda reply},}\ }\href@noop {} {\bibfield  {journal} {\bibinfo
		{journal} {Phys. Rev. Lett.}\ }\textbf {\bibinfo {volume} {104}},\ \bibinfo
	{pages} {198904} (\bibinfo {year} {2010})}\BibitemShut {NoStop}%
\bibitem [{\citenamefont {Cohen}\ and\ \citenamefont
	{Moerner}(2005)}]{cohen05b}%
\BibitemOpen
\bibfield  {author} {\bibinfo {author} {\bibfnamefont {A.~E.}\ \bibnamefont
		{Cohen}}\ and\ \bibinfo {author} {\bibfnamefont {W.~E.}\ \bibnamefont
		{Moerner}},\ }\bibfield  {title} {\enquote {\bibinfo {title} {Method for
			trapping and manipulating nanoscale objects in solution},}\ }\href@noop {}
{\bibfield  {journal} {\bibinfo  {journal} {App. Phys. Lett.}\ }\textbf
	{\bibinfo {volume} {86}},\ \bibinfo {eid} {093109} (\bibinfo {year}
	{2005})}\BibitemShut {NoStop}%
\bibitem [{\citenamefont {Cohen}(2005)}]{cohen05d}%
\BibitemOpen
\bibfield  {author} {\bibinfo {author} {\bibfnamefont {A.~E.}\ \bibnamefont
		{Cohen}},\ }\bibfield  {title} {\enquote {\bibinfo {title} {Control of
			nanoparticles with arbitrary two-dimensional force fields},}\ }\href@noop {}
{\bibfield  {journal} {\bibinfo  {journal} {Phys. Rev. Lett.}\ }\textbf
	{\bibinfo {volume} {94}},\ \bibinfo {pages} {118102} (\bibinfo {year}
	{2005})}\BibitemShut {NoStop}%
\bibitem [{\citenamefont {Jun}\ and\ \citenamefont {Bechhoefer}(2012)}]{jun12}%
\BibitemOpen
\bibfield  {author} {\bibinfo {author} {\bibfnamefont {Y.}~\bibnamefont
		{Jun}}\ and\ \bibinfo {author} {\bibfnamefont {J.}~\bibnamefont
		{Bechhoefer}},\ }\bibfield  {title} {\enquote {\bibinfo {title} {Virtual
			potentials for feedback traps},}\ }\href@noop {} {\bibfield  {journal}
	{\bibinfo  {journal} {Phys. Rev. E}\ }\textbf {\bibinfo {volume} {86}},\
	\bibinfo {pages} {061106} (\bibinfo {year} {2012})}\BibitemShut {NoStop}%
\bibitem [{\citenamefont {Gavrilov}\ \emph {et~al.}(2013)\citenamefont
	{Gavrilov}, \citenamefont {Jun},\ and\ \citenamefont
	{Bechhoefer}}]{gavrilov13}%
\BibitemOpen
\bibfield  {author} {\bibinfo {author} {\bibfnamefont {M.}~\bibnamefont
		{Gavrilov}}, \bibinfo {author} {\bibfnamefont {Y.}~\bibnamefont {Jun}}, \
	and\ \bibinfo {author} {\bibfnamefont {J.}~\bibnamefont {Bechhoefer}},\
}\bibfield  {title} {\enquote {\bibinfo {title} {Particle dynamics in a
		virtual harmonic potential},}\ }\href@noop {} {\bibfield  {journal} {\bibinfo
	{journal} {Proc. SPIE}\ }\textbf {\bibinfo {volume} {8810}} (\bibinfo {year}
{2013})}\BibitemShut {NoStop}%
\bibitem [{\citenamefont {Gavrilov}\ and\ \citenamefont
	{Bechhoefer}(2016)}]{Gavrilov16a}%
\BibitemOpen
\bibfield  {author} {\bibinfo {author} {\bibfnamefont {M.}~\bibnamefont
		{Gavrilov}}\ and\ \bibinfo {author} {\bibfnamefont {J.}~\bibnamefont
		{Bechhoefer}},\ }\bibfield  {title} {\enquote {\bibinfo {title} {Arbitrarily
			slow, non-quasistatic, isothermal transformations},}\ }\href@noop {}
{\bibfield  {journal} {\bibinfo  {journal} {EPL (Europhysics Letters)}\
	}\textbf {\bibinfo {volume} {114}},\ \bibinfo {pages} {50002} (\bibinfo
	{year} {2016})}\BibitemShut {NoStop}%
\bibitem [{\citenamefont {Gavrilov}\ \emph {et~al.}(2014)\citenamefont
	{Gavrilov}, \citenamefont {Jun},\ and\ \citenamefont
	{Bechhoefer}}]{gavrilov14}%
\BibitemOpen
\bibfield  {author} {\bibinfo {author} {\bibfnamefont {M.}~\bibnamefont
		{Gavrilov}}, \bibinfo {author} {\bibfnamefont {Y.}~\bibnamefont {Jun}}, \
	and\ \bibinfo {author} {\bibfnamefont {J.}~\bibnamefont {Bechhoefer}},\
}\bibfield  {title} {\enquote {\bibinfo {title} {Real-time calibration of a
		feedback trap},}\ }\href@noop {} {\bibfield  {journal} {\bibinfo  {journal}
	{Rev. Sci. Instrum.}\ }\textbf {\bibinfo {volume} {85}},\ \bibinfo {eid}
{095102} (\bibinfo {year} {2014})}\BibitemShut {NoStop}%
\bibitem [{\citenamefont {Gavrilov}\ \emph {et~al.}(2015)\citenamefont
	{Gavrilov}, \citenamefont {Koloczek},\ and\ \citenamefont
	{Bechhoefer}}]{gavrilov15}%
\BibitemOpen
\bibfield  {author} {\bibinfo {author} {\bibfnamefont {M.}~\bibnamefont
		{Gavrilov}}, \bibinfo {author} {\bibfnamefont {J.}~\bibnamefont {Koloczek}},
	\ and\ \bibinfo {author} {\bibfnamefont {J.}~\bibnamefont {Bechhoefer}},\
}\bibfield  {title} {\enquote {\bibinfo {title} {Feedback trap with
		scattering-based illumination},}\ }in\ \href@noop {} {\emph {\bibinfo
	{booktitle} {Novel Techniques in Microscopy}}}\ (\bibinfo  {publisher} {Opt.
Soc. Am.},\ \bibinfo {year} {2015})\ p.\ \bibinfo {pages} {JT3A.
4}\BibitemShut {NoStop}%
\bibitem [{\citenamefont {Cohen}\ and\ \citenamefont
	{Moerner}(2006)}]{cohen06a}%
\BibitemOpen
\bibfield  {author} {\bibinfo {author} {\bibfnamefont {A.~E.}\ \bibnamefont
		{Cohen}}\ and\ \bibinfo {author} {\bibfnamefont {W.~E.}\ \bibnamefont
		{Moerner}},\ }\bibfield  {title} {\enquote {\bibinfo {title} {Suppressing
			{B}rownian motion of individual biomolecules in solution},}\ }\href@noop {}
{\bibfield  {journal} {\bibinfo  {journal} {Proc. Natl. Acad. Sci. USA}\
	}\textbf {\bibinfo {volume} {103}},\ \bibinfo {pages} {4362--4365} (\bibinfo
	{year} {2006})}\BibitemShut {NoStop}%
\bibitem [{\citenamefont {Wang}\ and\ \citenamefont {Moerner}(2014)}]{wang14}%
\BibitemOpen
\bibfield  {author} {\bibinfo {author} {\bibfnamefont {Q.}~\bibnamefont
		{Wang}}\ and\ \bibinfo {author} {\bibfnamefont {W.~E.}\ \bibnamefont
		{Moerner}},\ }\bibfield  {title} {\enquote {\bibinfo {title} {Single-molecule
			motions enable direct visualization of biomolecular interactions in
			solution},}\ }\href@noop {} {\bibfield  {journal} {\bibinfo  {journal} {Nat.
			Methods}\ }\textbf {\bibinfo {volume} {11}},\ \bibinfo {pages} {556--558}
	(\bibinfo {year} {2014})}\BibitemShut {NoStop}%
\bibitem [{\citenamefont {Cohen}\ and\ \citenamefont
	{Moerner}(2007)}]{cohen07a}%
\BibitemOpen
\bibfield  {author} {\bibinfo {author} {\bibfnamefont {A.~E.}\ \bibnamefont
		{Cohen}}\ and\ \bibinfo {author} {\bibfnamefont {W.~E.}\ \bibnamefont
		{Moerner}},\ }\bibfield  {title} {\enquote {\bibinfo {title}
		{Principal-components analysis of shape fluctuations of single {DNA}
			molecules},}\ }\href@noop {} {\bibfield  {journal} {\bibinfo  {journal}
		{Proc. Natl. Acad. Sci. USA}\ }\textbf {\bibinfo {volume} {104}},\ \bibinfo
	{pages} {12622--12627} (\bibinfo {year} {2007})}\BibitemShut {NoStop}%
\bibitem [{\citenamefont {Goldsmith}\ and\ \citenamefont
	{Moerner}(2010)}]{Goldsmith2010}%
\BibitemOpen
\bibfield  {author} {\bibinfo {author} {\bibfnamefont {Randall~H.}\
		\bibnamefont {Goldsmith}}\ and\ \bibinfo {author} {\bibfnamefont {W.~E.}\
		\bibnamefont {Moerner}},\ }\bibfield  {title} {\enquote {\bibinfo {title}
		{Watching conformational- and photodynamics of single fluorescent proteins in
			solution},}\ }\href@noop {} {\bibfield  {journal} {\bibinfo  {journal}
		{Nature Chem.}\ }\textbf {\bibinfo {volume} {2}},\ \bibinfo {pages}
	{179--186} (\bibinfo {year} {2010})}\BibitemShut {NoStop}%
\bibitem [{\citenamefont {Fields}\ and\ \citenamefont
	{Cohen}(2011)}]{Fields2011}%
\BibitemOpen
\bibfield  {author} {\bibinfo {author} {\bibfnamefont {A.~P.}\ \bibnamefont
		{Fields}}\ and\ \bibinfo {author} {\bibfnamefont {A.~E.}\ \bibnamefont
		{Cohen}},\ }\bibfield  {title} {\enquote {\bibinfo {title} {Electrokinetic
			trapping at the one nanometer limit},}\ }\href@noop {} {\bibfield  {journal}
	{\bibinfo  {journal} {Proc. Natl. Acad. Sci. USA}\ }\textbf {\bibinfo
		{volume} {108}},\ \bibinfo {pages} {8937--8942} (\bibinfo {year}
	{2011})}\BibitemShut {NoStop}%
\bibitem [{\citenamefont {Germann}\ and\ \citenamefont
	{Davis}(2014)}]{germann14}%
\BibitemOpen
\bibfield  {author} {\bibinfo {author} {\bibfnamefont {J.~A.}\ \bibnamefont
		{Germann}}\ and\ \bibinfo {author} {\bibfnamefont {L.~M.}\ \bibnamefont
		{Davis}},\ }\bibfield  {title} {\enquote {\bibinfo {title} {Three-dimensional
			tracking of a single fluorescent nanoparticle using four-focus excitation in
			a confocal microscope},}\ }\href@noop {} {\bibfield  {journal} {\bibinfo
		{journal} {Opt. Express}\ }\textbf {\bibinfo {volume} {22}},\ \bibinfo
	{pages} {5641--5650} (\bibinfo {year} {2014})}\BibitemShut {NoStop}%
\bibitem [{\citenamefont {Kayci}\ \emph {et~al.}(2014)\citenamefont {Kayci},
	\citenamefont {Chang},\ and\ \citenamefont {Radenovic}}]{kayci14}%
\BibitemOpen
\bibfield  {author} {\bibinfo {author} {\bibfnamefont {M.}~\bibnamefont
		{Kayci}}, \bibinfo {author} {\bibfnamefont {H.-C.}\ \bibnamefont {Chang}}, \
	and\ \bibinfo {author} {\bibfnamefont {A.}~\bibnamefont {Radenovic}},\
}\bibfield  {title} {\enquote {\bibinfo {title} {Electron spin resonance of
		nitrogen-vacancy defects embedded in single nanodiamonds in an {ABEL}
		trap},}\ }\href@noop {} {\bibfield  {journal} {\bibinfo  {journal} {Nano
		Lett.}\ }\textbf {\bibinfo {volume} {14}},\ \bibinfo {pages} {5335--5341}
(\bibinfo {year} {2014})}\BibitemShut {NoStop}%
\bibitem [{\citenamefont {Lee}\ \emph {et~al.}(2015)\citenamefont {Lee},
	\citenamefont {Kwon},\ and\ \citenamefont {Pak}}]{lee15}%
\BibitemOpen
\bibfield  {author} {\bibinfo {author} {\bibfnamefont {D.~Y.}\ \bibnamefont
		{Lee}}, \bibinfo {author} {\bibfnamefont {C.}~\bibnamefont {Kwon}}, \ and\
	\bibinfo {author} {\bibfnamefont {H.~K.}\ \bibnamefont {Pak}},\ }\bibfield
{title} {\enquote {\bibinfo {title} {Nonequilibrium fluctuations for a
			single-particle analog of gas in a soft wall},}\ }\href@noop {} {\bibfield
	{journal} {\bibinfo  {journal} {Phys. Rev. Lett.}\ }\textbf {\bibinfo
		{volume} {114}},\ \bibinfo {pages} {060603} (\bibinfo {year}
	{2015})}\BibitemShut {NoStop}%
\bibitem [{Note1()}]{Note1}%
\BibitemOpen
\bibinfo {note} {See Supplemental Material [url], which includes Refs. \cite
	{Gavrilov16a,weigel14,gavrilov15,berglund08,gavrilov13,sekimoto97,sekimoto10,happel83,jun12,Jun14,gavrilov14,kawai07,parrondo09,roldan14,chiuchiu15,zulkowski14,dillenschneider09,callen85,KlagesBook13}}\BibitemShut
{NoStop}%
\bibitem [{\citenamefont {Callen}(1985)}]{callen85}%
\BibitemOpen
\bibfield  {author} {\bibinfo {author} {\bibfnamefont {H.~B.}\ \bibnamefont
		{Callen}},\ }\href@noop {} {\emph {\bibinfo {title} {Thermodynamics and an
			Introduction to Thermostatistics}}},\ \bibinfo {edition} {2nd}\ ed.\
(\bibinfo  {publisher} {Wiley},\ \bibinfo {year} {1985})\BibitemShut
{NoStop}%
\bibitem [{\citenamefont {Sekimoto}\ and\ \citenamefont
	{Sasa}(1997)}]{sekimoto97a}%
\BibitemOpen
\bibfield  {author} {\bibinfo {author} {\bibfnamefont {K.}~\bibnamefont
		{Sekimoto}}\ and\ \bibinfo {author} {\bibfnamefont {S.}~\bibnamefont
		{Sasa}},\ }\bibfield  {title} {\enquote {\bibinfo {title} {Complementarity
			relation for irreversible process derived from stochastic energetics},}\
}\href@noop {} {\bibfield  {journal} {\bibinfo  {journal} {J. Phys. Soc.
		Jap.}\ }\textbf {\bibinfo {volume} {66}},\ \bibinfo {pages} {3326--3328}
(\bibinfo {year} {1997})}\BibitemShut {NoStop}%
\bibitem [{\citenamefont {Schmiedl}\ and\ \citenamefont
	{Seifert}(2008)}]{schmiedl08}%
\BibitemOpen
\bibfield  {author} {\bibinfo {author} {\bibfnamefont {T.}~\bibnamefont
		{Schmiedl}}\ and\ \bibinfo {author} {\bibfnamefont {U.}~\bibnamefont
		{Seifert}},\ }\bibfield  {title} {\enquote {\bibinfo {title} {Efficiency at
			maximum power: An analytically solvable model for stochastic heat engines},}\
}\href@noop {} {\bibfield  {journal} {\bibinfo  {journal} {Euro. Phys.
		Lett.}\ } (\bibinfo {year} {2008})}\BibitemShut {NoStop}%
\bibitem [{\citenamefont {Maroney}(2005)}]{maroney05}%
\BibitemOpen
\bibfield  {author} {\bibinfo {author} {\bibfnamefont {O.~J.~E.}\
		\bibnamefont {Maroney}},\ }\bibfield  {title} {\enquote {\bibinfo {title}
		{The (absence of a) relationship between thermodynamic and logical
			reversibility},}\ }\href@noop {} {\bibfield  {journal} {\bibinfo  {journal}
		{Studies Hist. Phil. Mod. Phys.}\ }\textbf {\bibinfo {volume} {36}},\
	\bibinfo {pages} {355--374} (\bibinfo {year} {2005})}\BibitemShut {NoStop}%
\bibitem [{\citenamefont {Lopez-Suarez}\ \emph {et~al.}(2016)\citenamefont
	{Lopez-Suarez}, \citenamefont {Neri},\ and\ \citenamefont
	{Gammaitoni}}]{Lopez16}%
\BibitemOpen
\bibfield  {author} {\bibinfo {author} {\bibfnamefont {M.}~\bibnamefont
		{Lopez-Suarez}}, \bibinfo {author} {\bibfnamefont {I.}~\bibnamefont {Neri}},
	\ and\ \bibinfo {author} {\bibfnamefont {L.}~\bibnamefont {Gammaitoni}},\
}\bibfield  {title} {\enquote {\bibinfo {title} {Sub-{$k_BT$}
		micro-electromechanical irreversible logic gate},}\ }\href@noop {} {\bibfield
{journal} {\bibinfo  {journal} {Nat Commun}\ }\textbf {\bibinfo {volume}
	{7}} (\bibinfo {year} {2016})}\BibitemShut {NoStop}%
\bibitem [{\citenamefont {Maroney}(2009)}]{maroney09}%
\BibitemOpen
\bibfield  {author} {\bibinfo {author} {\bibfnamefont {O.~J.~E.}\
		\bibnamefont {Maroney}},\ }\bibfield  {title} {\enquote {\bibinfo {title}
		{Generalizing {L}andauer's principle},}\ }\href@noop {} {\bibfield  {journal}
	{\bibinfo  {journal} {Phy. Rev. E}\ }\textbf {\bibinfo {volume} {79}},\
	\bibinfo {pages} {031105} (\bibinfo {year} {2009})}\BibitemShut {NoStop}%
\bibitem [{\citenamefont {Rold{\'a}n}\ \emph {et~al.}(2015)\citenamefont
	{Rold{\'a}n}, \citenamefont {Neri}, \citenamefont {D{\"o}rpinghaus},
	\citenamefont {Meyr},\ and\ \citenamefont {J{\"u}licher}}]{roldan15}%
\BibitemOpen
\bibfield  {author} {\bibinfo {author} {\bibfnamefont {E.}~\bibnamefont
		{Rold{\'a}n}}, \bibinfo {author} {\bibfnamefont {I.}~\bibnamefont {Neri}},
	\bibinfo {author} {\bibfnamefont {M.}~\bibnamefont {D{\"o}rpinghaus}},
	\bibinfo {author} {\bibfnamefont {H.}~\bibnamefont {Meyr}}, \ and\ \bibinfo
	{author} {\bibfnamefont {F.}~\bibnamefont {J{\"u}licher}},\ }\bibfield
{title} {\enquote {\bibinfo {title} {Decision making in the arrow of time},}\
}\href@noop {} {\bibfield  {journal} {\bibinfo  {journal} {Phys. Rev. Lett.}\
}\textbf {\bibinfo {volume} {115}},\ \bibinfo {pages} {250602} (\bibinfo
{year} {2015})}\BibitemShut {NoStop}%
\bibitem [{\citenamefont {Dillenschneider}\ and\ \citenamefont
	{Lutz}(2009)}]{dillenschneider09}%
\BibitemOpen
\bibfield  {author} {\bibinfo {author} {\bibfnamefont {Raoul}\ \bibnamefont
		{Dillenschneider}}\ and\ \bibinfo {author} {\bibfnamefont {Eric}\
		\bibnamefont {Lutz}},\ }\bibfield  {title} {\enquote {\bibinfo {title}
		{Memory erasure in small systems},}\ }\href@noop {} {\bibfield  {journal}
	{\bibinfo  {journal} {Phys. Rev. Lett.}\ }\textbf {\bibinfo {volume} {102}},\
	\bibinfo {pages} {210601} (\bibinfo {year} {2009})}\BibitemShut {NoStop}%
\bibitem [{\citenamefont {Weigel}\ \emph {et~al.}(2014)\citenamefont {Weigel},
	\citenamefont {Sebesta},\ and\ \citenamefont {Kukura}}]{weigel14}%
\BibitemOpen
\bibfield  {author} {\bibinfo {author} {\bibfnamefont {A.}~\bibnamefont
		{Weigel}}, \bibinfo {author} {\bibfnamefont {A.}~\bibnamefont {Sebesta}}, \
	and\ \bibinfo {author} {\bibfnamefont {P.}~\bibnamefont {Kukura}},\
}\bibfield  {title} {\enquote {\bibinfo {title} {Dark field microspectroscopy
		with single molecule fluorescence sensitivity},}\ }\href@noop {} {\bibfield
{journal} {\bibinfo  {journal} {ACS Photonics}\ }\textbf {\bibinfo {volume}
	{1}},\ \bibinfo {pages} {848--856} (\bibinfo {year} {2014})}\BibitemShut
{NoStop}%
\bibitem [{\citenamefont {Berglund}\ \emph {et~al.}(2008)\citenamefont
	{Berglund}, \citenamefont {McMahon}, \citenamefont {McClelland},\ and\
	\citenamefont {Liddle}}]{berglund08}%
\BibitemOpen
\bibfield  {author} {\bibinfo {author} {\bibfnamefont {A.~J.}\ \bibnamefont
		{Berglund}}, \bibinfo {author} {\bibfnamefont {M.~D.}\ \bibnamefont
		{McMahon}}, \bibinfo {author} {\bibfnamefont {J.~J.}\ \bibnamefont
		{McClelland}}, \ and\ \bibinfo {author} {\bibfnamefont {J.~A.}\ \bibnamefont
		{Liddle}},\ }\bibfield  {title} {\enquote {\bibinfo {title} {Fast, bias-free
			algorithm for tracking single particles with variable size and shape},}\
}\href@noop {} {\bibfield  {journal} {\bibinfo  {journal} {Opt. Express}\
}\textbf {\bibinfo {volume} {16}},\ \bibinfo {pages} {14064--14075} (\bibinfo
{year} {2008})}\BibitemShut {NoStop}%
\bibitem [{\citenamefont {Sekimoto}(1997)}]{sekimoto97}%
\BibitemOpen
\bibfield  {author} {\bibinfo {author} {\bibfnamefont {K.}~\bibnamefont
		{Sekimoto}},\ }\bibfield  {title} {\enquote {\bibinfo {title} {Kinetic
			characterization of heat bath and the energetics of thermal ratchet
			models},}\ }\href@noop {} {\bibfield  {journal} {\bibinfo  {journal} {J.
			Phys. Soc. Jap.}\ }\textbf {\bibinfo {volume} {66}},\ \bibinfo {pages}
	{1234--1237} (\bibinfo {year} {1997})}\BibitemShut {NoStop}%
\bibitem [{\citenamefont {Happel}\ and\ \citenamefont
	{Brenner}(1983)}]{happel83}%
\BibitemOpen
\bibfield  {author} {\bibinfo {author} {\bibfnamefont {J.}~\bibnamefont
		{Happel}}\ and\ \bibinfo {author} {\bibfnamefont {H.}~\bibnamefont
		{Brenner}},\ }\href@noop {} {\emph {\bibinfo {title} {Low Reynolds Number
			Hydrodynamics: With Special Applications to Particulate Media}}}\ (\bibinfo
{publisher} {Martinus Nijhoff},\ \bibinfo {year} {1983})\BibitemShut
{NoStop}%
\bibitem [{\citenamefont {Kawai}\ \emph {et~al.}(2007)\citenamefont {Kawai},
	\citenamefont {Parrondo},\ and\ \citenamefont {{V}an~den Broeck}}]{kawai07}%
\BibitemOpen
\bibfield  {author} {\bibinfo {author} {\bibfnamefont {R.}~\bibnamefont
		{Kawai}}, \bibinfo {author} {\bibfnamefont {J.~M.~R.}\ \bibnamefont
		{Parrondo}}, \ and\ \bibinfo {author} {\bibfnamefont {C.}~\bibnamefont
		{{V}an~den Broeck}},\ }\bibfield  {title} {\enquote {\bibinfo {title}
		{Dissipation: The phase-space perspective},}\ }\href@noop {} {\bibfield
	{journal} {\bibinfo  {journal} {Phys. Rev. Lett.}\ }\textbf {\bibinfo
		{volume} {98}},\ \bibinfo {pages} {080602} (\bibinfo {year}
	{2007})}\BibitemShut {NoStop}%
\bibitem [{\citenamefont {Parrondo}\ \emph {et~al.}(2009)\citenamefont
	{Parrondo}, \citenamefont {{V}an~den Broeck},\ and\ \citenamefont
	{Kawai}}]{parrondo09}%
\BibitemOpen
\bibfield  {author} {\bibinfo {author} {\bibfnamefont {J.~M.~R.}\
		\bibnamefont {Parrondo}}, \bibinfo {author} {\bibfnamefont {C.}~\bibnamefont
		{{V}an~den Broeck}}, \ and\ \bibinfo {author} {\bibfnamefont
		{R.}~\bibnamefont {Kawai}},\ }\bibfield  {title} {\enquote {\bibinfo {title}
		{Entropy production and the arrow of time},}\ }\href@noop {} {\bibfield
	{journal} {\bibinfo  {journal} {New J. Phys.}\ }\textbf {\bibinfo {volume}
		{11}},\ \bibinfo {pages} {073008} (\bibinfo {year} {2009})}\BibitemShut
{NoStop}%
\bibitem [{\citenamefont {Rold{\'a}n}\ \emph {et~al.}(2014)\citenamefont
	{Rold{\'a}n}, \citenamefont {Mart{\'\i}nez}, \citenamefont {Parrondo},\ and\
	\citenamefont {Petrov}}]{roldan14}%
\BibitemOpen
\bibfield  {author} {\bibinfo {author} {\bibfnamefont {{\'E}.}~\bibnamefont
		{Rold{\'a}n}}, \bibinfo {author} {\bibfnamefont {I.~A.}\ \bibnamefont
		{Mart{\'\i}nez}}, \bibinfo {author} {\bibfnamefont {J.~M.~R.}\ \bibnamefont
		{Parrondo}}, \ and\ \bibinfo {author} {\bibfnamefont {D.}~\bibnamefont
		{Petrov}},\ }\bibfield  {title} {\enquote {\bibinfo {title} {Universal
			features in the energetics of symmetry breaking},}\ }\href@noop {} {\bibfield
	{journal} {\bibinfo  {journal} {Nature Phys.}\ }\textbf {\bibinfo {volume}
		{10}},\ \bibinfo {pages} {457--461} (\bibinfo {year} {2014})}\BibitemShut
{NoStop}%
\bibitem [{\citenamefont {Chiuchi{\'u}}(2015)}]{chiuchiu15}%
\BibitemOpen
\bibfield  {author} {\bibinfo {author} {\bibfnamefont {D.}~\bibnamefont
		{Chiuchi{\'u}}},\ }\bibfield  {title} {\enquote {\bibinfo {title}
		{Time-dependent study of bit reset},}\ }\href@noop {} {\bibfield  {journal}
	{\bibinfo  {journal} {EPL (Europhysics Letters)}\ }\textbf {\bibinfo {volume}
		{109}},\ \bibinfo {pages} {30002} (\bibinfo {year} {2015})}\BibitemShut
{NoStop}%
\bibitem [{\citenamefont {Zulkowski}\ and\ \citenamefont
	{DeWeese}(2014)}]{zulkowski14}%
\BibitemOpen
\bibfield  {author} {\bibinfo {author} {\bibfnamefont {Patrick~R.}\
		\bibnamefont {Zulkowski}}\ and\ \bibinfo {author} {\bibfnamefont
		{Michael~R.}\ \bibnamefont {DeWeese}},\ }\bibfield  {title} {\enquote
	{\bibinfo {title} {Optimal finite-time erasure of a classical bit},}\
}\href@noop {} {\bibfield  {journal} {\bibinfo  {journal} {Phys. Rev. E}\
}\textbf {\bibinfo {volume} {89}},\ \bibinfo {pages} {052140} (\bibinfo
{year} {2014})}\BibitemShut {NoStop}%
\end{thebibliography}

%

\clearpage
\setboolean{@twoside}{false}


\section*{Supplementary material}

# Supplemental Material for "Erasure without work in an asymmetric, double-well potential"

Momčilo Gavrilov and John Bechhoefer*
*Department of Physics, Simon Fraser University, Burnaby, B.C., V5A 1S6, Canada*

## EXPERIMENTAL SETUP

We give more details on the experimental setup, which is similar to that described in [1]. For experimental tests here, we collected about two weeks worth of data. We trap an overdamped silica bead of diameter 1.5 μm in a virtual potential created by a feedback loop with update time $\Delta t = 5$ ms (Fig. S1, from Ref. [1]). The apparatus is based on a home-built dark-field microscope equipped with a 60x Olympus NA=0.95 air objective [2, 3]. An LED source (660 nm, Thorlabs) illuminates the sample, and a camera (Andor iXon DV-885) takes a $50 \times 20$ pixel image every $\Delta t = 5$ ms, with an exposure $t_c = 0.5$ ms. The potential is a double well along the $x$ axis and harmonic along the $y$ axis. A LabVIEW program processes acquired images by thresholding the background and applying a modified centroid algorithm to a small region of interest [4, 5].

Based on the estimated position and the virtual potential, voltages are applied at a time $t_d = 5$ ms after the middle of the exposure. The DAQ voltages are amplified 15x using a custom-built amplifier and held constant for 5 ms, until the next update. The working particles in our trap were chosen to be large enough and heavy enough that gravity confines their motion to within the depth of focus of our microscope, yet small enough to diffuse freely in the horizontal plane. In particular, the particles had a diameter $2a = 1.5$ μm and were made of silica (density $\rho_s = 2.2$ gm/cc, in water).

## TRAPPING IN THE $XY$-PLANE

We measure the dynamics of a particle trapped in a time-dependent, double-well, virtual potential. The total potential includes a harmonic contribution in the $y$ component, $\frac{1}{2}\kappa y^2$, which is kept constant throughout the experiments. Since it plays no role in work calculations, it is ignored in the rest of the analysis, which focuses on the one-dimensional motion along $x$.

In one dimension, the double-well potential is

$$U(x,t) = 4E_b \left[ -\tfrac{1}{2} g(t)\, \tilde{x}^2 + \tfrac{1}{4} \tilde{x}^4 - A f(t)\, \tilde{x} \right], \quad \text{(S1)}$$

where the scaled coordinate $\tilde{x}(x,t)$ has different values

---

* email: johnb@sfu.ca

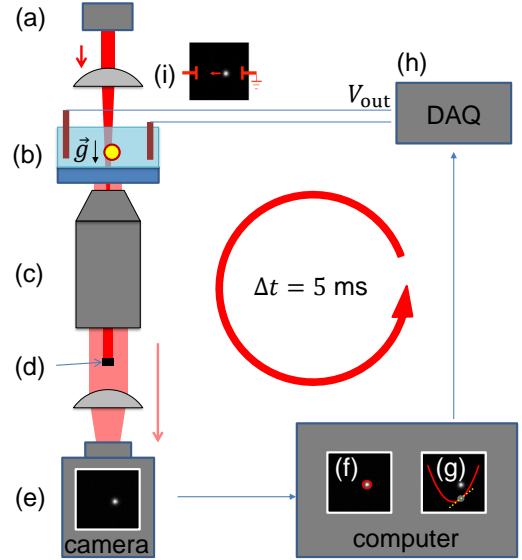

FIG. S1. (Color online) Schematic of feedback trap operation. (a) LED light source illuminates the trapped silica bead. (b) Bead in sample cell sinks under gravity and diffuses predominantly in the XY plane. (c) High-NA microscope objective collects light scattered from bead and also directly from LED source. (d) Beam blocker stops the LED beam, allowing only scattered light from bead to reach the camera. (e) Camera takes image. (f) Image-processing program estimates bead position and (g) calculates the force based by the imposed potential. (h) DAQ applies voltage proportional to calculated force to electrodes. (i) Forces due to electric field and thermal fluctuations move bead to a new position. The feedback loop repeats indefinitely, with cycle time $\Delta t = 5$ ms. Figure from Ref. [1].

on the left and right sides of the potential,

$$\tilde{x}(x,t) = \begin{cases} \frac{x}{x_0} & x < 0 \\ \frac{x}{x_0} \left[1 + m(t)(\eta - 1)\right]^{-1} & x \geq 0, \end{cases} \quad \text{(S2)}$$

and controls the asymmetry of the double-well potential. The tilt scale is set by $|A|/kT = 0.25$; the sign determines the well used for erasure.

The dimensionless functions $g(t)$, $f(t)$, and $m(t)$ together define the experimental protocol and smoothly vary between 0 and 1. See Fig. S5 for explicit forms. The function $g(t)$ controls the barrier height, $f(t)$ tilts the potential to favor the larger well ($A > 0$) or the smaller well ($A < 0$), and $m(t)$ controls the asymmetry ratio $\eta$.

## TRAPPING IN THE Z-DIRECTION

A combination of repulsive electrostatic and attractive gravitational forces confines the particle in the $z$-direction (Fig.S1b). The confinement is thus by physical potentials, as opposed to the virtual potentials used to confine in the $y$-direction and to create the features of interest along the $x$-direction. The decoupling that justifies neglecting $y$-effects also applies to the $z$-direction.

However, there is one important new feature that arises from the $z$-motion, which is that the lateral ($x$) diffusion constant, $D_x$ depends on the height ($z$) and thus fluctuates. The experimental results here measure the work by a changing potential. Since the $z$-dependence of $U(x, y, z, t)$ is constant in time, it makes no direct contribution. But the value of $D_x$ enters implicitly linearly into the calculation of work done by changes along $x$. Fortunately, the dependence in $D_x$ is linear, so that the fluctuations in $D_x$ lead to similar fluctuations in the stochastic work $W$. Since, in any case, we measure an average $D$, the fact that such a value is fluctuating will not change the average value of $W$. It will increase the variance, but, as we argue below, even that is negligible.

We can determine the time scale of z-fluctuations $\tau_z$ (and, hence, $D$ fluctuations) from the autocorrelation of intensity measurements of the bead, as intensity is a function of bead height $z$. We measure $\approx 0.3$ s. This is to be compared with the total measurement time per cycle, compiled over many cycles of 30 min = 1800 s.

Finally, the values that we calculate in settings where we feel we know precisely what to expect (e.g., the $kT \ln 2$ work to erase a symmetric memory) are consistent with measurements based only on $x$-trajectories.

## STOCHASTIC WORK ESTIMATE

During the protocol, we record the shape of the potential $U(x, t_n)$ and the observed position $\bar{x}_n$ at all time steps $t_n$, from which we estimate work by discretizing Sekimoto's formula [6, 7] for the stochastic work. $W_{\text{cyc}} = \int_0^{t_{\text{cyc}}} dt \left. \frac{\partial U(x,t)}{\partial t} \right|_{x=x(t)}$, where $t_{\text{cyc}}$ is the duration of erasure cycle in seconds. To report work values, we scale $t_{\text{cyc}}$ by $\tau_0 = [(1 + \eta)x_0]^2/D$, which is the time needed for a particle to freely diffuse the distance $(1 + \eta)x_0$. The dimensionless cycle time $\tau \equiv t_{\text{cyc}}/\tau_0$. We collect work values for fixed cycle time $\tau$ for 30 min., after which we change $\tau$ and repeat the measurements.

## EXPERIMENTAL PARAMETERS

For a 1.5-$\mu$m-diameter bead, typical mobility values in our sample cell are $\mu \approx 0.1$ $\mu$m/V/s. The diffusivity $D \approx$ 0.23 $\mu$m$^2$/s, which is 0.67 times the Stokes-Einstein value of 0.35 $\mu$m$^2$/s for a similar sphere far from boundaries and consistent with the increased drag expected for a sphere near a boundary [8].

For a virtual potential, the feedback loop update time $\Delta t$ should be fast compared to the (overdamped) local relaxation time $t_r$ of particles fluctuating about the local potential minimum at $\pm x_0$ [9]. To satisfy this constraint, we select $t_r > \Delta t/0.2$ and further choose

$$x_0^2 = 8(E_b/kT)Dt_r > 8(E_b/kT)D\Delta t/0.2, \quad \text{(S3)}$$

where $E_b$ is the barrier height [10]. Since the barrier-hop time for the potential in Eq. 1 is

$$\tau_{\text{hop}} = \left(\frac{\sqrt{2}\pi}{16}\right)\left(\frac{\exp(E_b/kT)}{E_b/kT}\right), \quad \text{(S4)}$$

the energy barrier $E_b/kT = 13$ imposes a barrier-hop time $\tau_{\text{hop}} \approx 10^5$. By contrast, the longest dimensionless cycle time was $\tau = 50$. Using the values for $D$ and $E_b$, we set $x_0 = 0.77$ $\mu$m.

## DISCRETIZATION EFFECTS

In this section, we derive the correction to the apparent value of the average work that we measure, arising from finite-time discretization effects present in a feedback trap. In the main article, we decomposed the work to erase an asymmetric memory into three components:

- compression of one well;
- erasure protocol for a symmetric potential;
- expansion of one well.

We showed that the work for cyclic transformations of symmetric virtual potential is not affected by the finite-time discretization effect of a feedback trap [10]. Thus, the feedback trap can estimate the average work to erase a symmetric memory with high precision. However, if the transformation of a potential is not cyclic, the prediction of work done by the feedback trap requires finite-time corrections. In an earlier work [9], it was shown how finite update time $\Delta t$, finite camera exposure $t_c$, delay $t_d$, and observational noise $\chi$ affect work needed to change the stiffness of a harmonic potential that traps a single bead.

The mean work needed to change the stiffness $\kappa$ of a harmonic potential $U(x, t) = \frac{1}{2}\kappa(t)x^2$ from $\kappa_i$ to $\kappa_f$ is:

$$\begin{aligned} \frac{\langle W \rangle_{\text{har}}}{kT} &= \frac{1}{kT}\left\langle \int dt\, \frac{\partial U}{\partial t} \right\rangle \\ &= \frac{1}{2kT}\int dt\, \langle \dot{\kappa}(t)\, x^2 \rangle \\ &= \frac{1}{2}\int dt\, \dot{\kappa}(t)\frac{\langle x^2 \rangle}{kT} \end{aligned} \quad \text{(S5)}$$

For a particle in a physical harmonic potential with stiffness $\kappa^{(c)}$, the expression $\langle x^2 \rangle = \frac{kT}{\kappa^{(c)}}$ holds, so the work done by physical continuous potential $\langle W \rangle^{(c)}_{\text{har}}$ is

$$\frac{\langle W \rangle^c_{\text{har}}}{kT} = \frac{1}{2}\int dt\, \dot{\kappa}^{(c)} \frac{\langle x^2 \rangle}{kT} = \frac{1}{2}\int dt\, \frac{\dot{\kappa}^{(c)}}{\kappa^{(c)}}$$
$$= \frac{1}{2}\int \frac{d\kappa^{(c)}}{\kappa^{(c)}} = \frac{1}{2}\ln\frac{\kappa_f}{\kappa_i}\,. \quad \text{(S6)}$$

For a virtual harmonic potential, $\langle x^2 \rangle = \frac{kT}{\kappa^{(c)}}$ is generally not accurate [9]. It is more convenient to describe such a system by a feedback gain $\alpha$. The conversion between the stiffness $k$ of the trap and the feedback gain is given by $\alpha = \kappa D \Delta t$. For the feedback trap with delay $t_d = \Delta t$, the work done when transforming the feedback gain from $\alpha_i$ to $\alpha_f$ is given by [9]

$$\frac{\langle W \rangle^d_{\text{har}}}{kT} = \frac{1}{2}\ln\left(\frac{\alpha_f}{\alpha_i}\right) + \underbrace{\frac{1}{2}\ln\left[\left(\frac{2+\alpha_f}{2+\alpha_i}\right)^{\frac{1}{3}}\left(\frac{1-\alpha_f}{1-\alpha_i}\right)^{\frac{4}{3}}\right]}_{\text{finite-time correction}}$$
$$= \frac{\langle W \rangle^{(c)}_{\text{har}}}{kT} + \frac{\langle W \rangle_{\text{cor}}}{kT}\,. \quad \text{(S7)}$$

The feedback gain $\alpha$ can also be interpreted as the ratio between the update time $\Delta t$ and the relaxation time $t_r$ of a particle within the basin of a harmonic trap $\alpha = \Delta t/t_r$. For a feedback trap with instantaneous response $\Delta t \to 0$, the correction term

$$\frac{\langle W \rangle_{\text{cor}}}{kT} = \frac{1}{2}\ln\left[\left(\frac{2+\alpha_f}{2+\alpha_i}\right)^{\frac{1}{3}}\left(\frac{1-\alpha_f}{1-\alpha_i}\right)^{\frac{4}{3}}\right] \quad \text{(S8)}$$

converges to 0, and the work done by the virtual potential approaches the result for the continuous case.

Here, we are interested in estimating the work to change the size of one well of a double-well potential. Such a change is illustrated in Fig. S2, where one side of a double-well potential is expanded. To calculate the work for such change, we first approximate the bottom of a double-well with the harmonic trap of the same maximal curvature. The maximal curvature $|\kappa_m|$ is found as the second derivative of a potential $U(x)$. For the double-well potential in Eq. S1, which is also plotted as the dotted line in Fig. S2, the curvature around the minimum is

$$|\kappa_m| = \left|\frac{d^2 U(x)}{dx^2}\right|_{x=x_m} = 8\frac{E_b}{(\eta x_0)^2}\,.$$

We can now approximate the double-well with a harmonic potential,

$$U(x) = \frac{1}{2}\kappa x^2 = \frac{1}{2}|\kappa_m|x^2 = 4\frac{E_b}{(\eta x_0)^2}(x \pm \eta x_0)^2\,.$$

This harmonic approximation is shown in blue in Fig. S2. Note that the particle spends more than 99% of its time

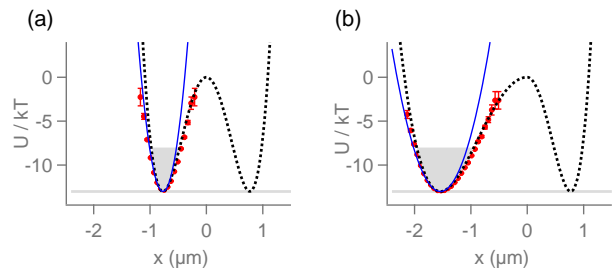

FIG. S2. (Color online) Change in symmetry of a double-well potential. a) Symmetric double-well potential ($\eta = 1$). b) Asymmetric double-well potential. The left well is twice as large as the right well ($\eta = 2$). Dotted lines show parametrizations of the double-well potentials. The blue line indicates the harmonic approximation to the double well potential, with matched curvatures. The time series of particle positions is used to estimate the Boltzmann distribution $p(\tilde{x})$ and the reconstruction of potential $U(\tilde{x}) = U_0 - \ln(p(\tilde{x}))$, which is shown by red markers. Shaded area indicates region where particle spends more than 99% of its time in the well.

in the shaded area, where the difference between harmonic and double-well potentials is small.

For the harmonic approximation, we calculate the feedback gain

$$\alpha = \kappa D \Delta t = \frac{8 E_b D \Delta t}{(\eta x_0)^2}\,,$$

and we apply Eq. S7 to find work corrections due to the finite time scales of a feedback trap. In the example shown in Fig. S2, we change $\eta$ from $\eta_i = 1$ to $\eta_f = 2$. This operation doubles the phase space available to the particle. For the continuous potential, the required work is $\langle W \rangle^c_{\text{har}}/kT = -\ln 2 \approx -0.69$. Such a change is done with a single harmonic trap and should not be confused with the work needed to erase memory (Landauer limit). For $\eta_i = 1$, the feedback gain is $\alpha_i = 0.2$, and for $\eta_f = 2$, the gain is $\alpha_f = 0.05$. Therefore, the expected work done by a virtual potential is $\langle W \rangle^d_{\text{har}}/kT = -0.82$. We can then estimate the difference between work done by the virtual and real potentials:

$$\frac{\left|\langle W \rangle^d_{\text{har}} - \langle W \rangle^c_{\text{har}}\right|}{kT} \approx 0.82 - 0.69 = 0.13\,.$$

Finally, we test this analysis with our experimental data, using the segments of cycles where the expansion of double-well potential occurs. We isolate data during which one well of the double-well potential is expanded by factor of $\eta = 2$ and also compressed by the same factor. In Fig. S3, we plot as a function of inverse cycle time $\tau^{-1}$ the mean work to expand (in blue), work to compress (in red), and work to do a cyclic operation of expanding and compressing (in gray) one well of the double-well potential. In the arbitrarily slow limit,

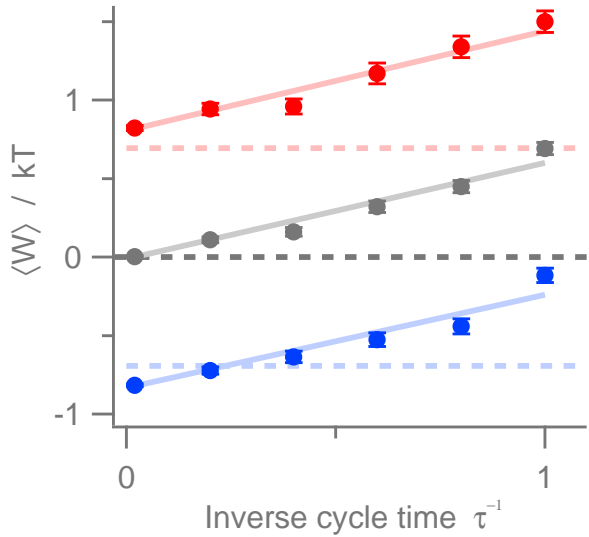

FIG. S3. (Color online) Work to change the size of one well of a double-well potential by factor ($\eta = 2$) for expansion (blue), compression (red), and both expansion and compression (gray). Experimental data are shown by markers; solid lines are linear fits used to estimate work in an arbitrarily slow limit, which corresponds to the y-axis intercept. Dotted lines are expected work values for the continuous potential.

we measure $\langle W \rangle^d / kT = -0.83 \pm 0.01$ for potential expansion, $\langle W \rangle^d / kT = 0.80 \pm 0.02$ for compression and $\langle W \rangle^d / kT = -0.013 \pm 0.007$ for the cyclic operation of expanding and compressing the potential. Operations of changing the stiffness, either just expansion or just compression, show different work values than predicted for the continuous potential. For cyclic operations such as expanding and then compressing, these corrections cancel and the measured work is exact.

## WORK TO ERASE ASYMMETRIC ONE-BIT MEMORY IN A VIRTUAL POTENTIAL

In this section, we derive the discretized version of Eq. 1 from the main article. Again, the work is split into three parts: compression, symmetric erasure, and expansion. This last either happens in all cases or never happens. The work to expand or compress the potential by a factor $\eta$ is obtained from Eq. S7, by expressing the feedback gain $\alpha$ as function of asymmetry $\alpha(\eta) = \kappa D \Delta t = \frac{8 E_b}{(\eta x_0)^2} D \Delta t$ and further substituting $\alpha_i = \alpha(1)$ and $\alpha_f = \alpha(\eta)$.

$$\frac{\langle W \rangle^d}{kT} = \ln 2 \pm \frac{1}{2} \ln \eta \pm \frac{1}{2} \frac{\langle W \rangle_{\text{cor}}}{kT} \qquad (S9)$$

The first two terms in work done by the virtual potential are the same as for the physical potential. The last term, $\langle W \rangle_{\text{cor}}$, corrects the finite-time discretization effects defined in Eq. S7. The correction depends on several parameters: $E_b$, $\eta$, $x_0$, $\Delta t$, and $D$. All except the diffusivity $D$ are controllable. The diffusivity is a property of a trapped bead, and external factors such as the geometric shape of the trapping region, temperature, and fluid viscosity affect its value. We use several nominally identical beads for measurements. The diffusivity among beads varies up to 5%, but we can measure $D$ to high precision [11] and properly select the cycle time for each measurement. This variation in diffusivity also affects the feedback gain $\alpha$ by 5% and introduces second-order work corrections. Since they are less than 0.007 $kT$, we do not consider them in our work prediction. In Fig. 2b, we plot Eq. S9 using the mean diffusivity $D = 0.23~\mu\text{m}^2/\text{s}$.

## WORK FOR THERMODYNAMICALLY IRREVERSIBLE ERASURE

The minimal work to erase memory in the arbitrarily slow limit can be achieved only if the erasure protocol is thermodynamically reversible. In the main article, we showed an example of erasure protocol that is thermodynamically reversible (Fig. 1) and one that is thermodynamically irreversible (Fig. 3). The irreversible protocol erases the asymmetric memory by first lowering down the barrier, tilting the potential, rising the barrier and untilting. The step of lowering down the barrier and mixing two nonequilibrium states is irreversible; as a consequence, when the entire protocol is played backward in time, it will not take the system to the same initial nonequilibrium state (see Fig. 3b). The irreversibility of the protocol generates extra entropy. Here, we calculate the work due to this entropy production.

We start from a general relationship for a protocol where the probabilities of forward and backward trajectories differ [1, 12–15]:

$$\frac{\langle W \rangle}{kT} \geq H - \frac{\Delta F}{kT} + D_{KL}(p_i || \bar{p}_i). \qquad (S10)$$

The inequality applies to work done at finite times, and it becomes an equality in the arbitrarily slow limit. The change in free energy and information between initial and final states is given by the nonequilibrium free energy difference $H - \Delta F / kT$. Here, in the initial state, the memory has a Shannon entropy of $H = \ln 2$. In the final state, after erasure, the Shannon entropy is zero, since the system occupies one state with certainty.

To calculate the change in free energy of the states, $\Delta F$, let $F_k$ be the free energy of state $k$ before erasure and $p_k$ the probability that the system is initially in that state. After the erasure, the system is in the standard

state, whose free energy is denoted $F_0$. Then [15],

$$\Delta F = \underbrace{\sum p_k F_k}_{\text{initial}} - \underbrace{F_0}_{\text{final}} . \tag{S11}$$

For a symmetric memory, $F_k = F_0$, and $\Delta F = 0$.

For an asymmetric memory, the two states, $L$ and $R$, initially have $p_L = p_R = \frac{1}{2}$, and

$$\begin{aligned}\frac{\Delta F}{kT} &= \frac{1}{kT}\left[\underbrace{\tfrac{1}{2}(F_L + F_R)}_{\text{initial}} - \underbrace{F_{R,L}}_{\text{final}}\right] \\ &= \mp \frac{1}{kT}\left[\tfrac{1}{2}(F_L - F_R)\right] = \mp \tfrac{1}{2}\ln\eta\,.\end{aligned} \tag{S12}$$

The last term in Eq. S10 quantifies irreversibility, using the Kullback-Leibler (KL) divergence (relative entropy),

$$D_{KL}(p_i || \bar{p}_i) = \sum p_i \ln \frac{p_i}{\bar{p}_i}\,,$$

between the probability of the initial state $p_i$ and the probability of a system to return to the same state when time is reversed, $\bar{p}_i$. The KL divergence does not depend on the final state of a system. The initial probabilities are $p_L = p_R = 0.5$ for both protocols. When the first protocol is reversed, it will lead to probabilities $\bar{p}_L = \bar{p}_R = 0.5$, and $D_{KL}$ is zero. But when the second is reversed, the system will end up with probabilities [1]

$$\bar{p}_L = \frac{1}{1+\eta}, \qquad \bar{p}_R = \frac{\eta}{1+\eta}, \tag{S13}$$

and

$$\begin{aligned}D_{KL}(p_i || \bar{p}_i) &= \sum_{i=L,R} p_i \ln \frac{p_i}{\bar{p}_i} \\ &= \frac{1}{2}\left[\ln\left(\frac{1/2}{1/(\eta+1)}\right) + \ln\left(\frac{1/2}{\eta/(\eta+1)}\right)\right] \\ &= \ln\left(\frac{\eta+1}{2\sqrt{\eta}}\right).\end{aligned} \tag{S14}$$

Finally, let a memory be represented by an asymmetric double-well potential, where the particle has an equal probability to be in either well $p_L = p_R = 0.5$, and let the ratio of well sizes be given by $\eta$. Then the work needed to erase such a memory by lowering down the barrier, tilting the potential, raising the barrier, and untilting is given by

$$\frac{\langle W \rangle}{kT} = \underbrace{\ln 2}_{\text{macrostates}} \underbrace{\pm \frac{1}{2}\ln\eta}_{\text{microstates}} \underbrace{\pm \frac{1}{2}\frac{\langle W \rangle_{\text{cor}}}{kT}}_{\text{finite-time effects}} \underbrace{+ \ln\left(\frac{\eta+1}{2\sqrt{\eta}}\right)}_{\text{thermodynamic irreversibility}}. \tag{S15}$$

This formula includes work contributions from the reduction in information-bearing degrees of freedom (macrostates), changes in non-information-bearing degrees of freedom (micro states), and corrections due to both finite-time effects and the irreversibility of the erasure protocol. This function is plotted in Fig. 4.

## THE ERASURE PROTOCOL

In this section, we define precisely the erasure protocols used for asymmetric bit erasure. The erasure protocol separates explicitly the operations of changing the potential symmetry and erasing a bit with a fixed asymmetry. The function $m(t)$ in Eq. S1 controls the asymmetry, while the functions $g(t)$ and $f(t)$ are responsible for the erasure of a bit with a fixed symmetry. In the reversible protocol, changing the asymmetry is never done while erasing the memory. Figure S5 shows the protocols for the erasure of the asymmetric bit. For the thermodynamically reversible protocol, the size of the bigger well is linearly decreased, making it symmetric when time reaches $\frac{1}{4}\tau$ (see S5a, blue line). Once the potential becomes symmetric, we erase the symmetric bit by lowering the barrier using the quadratic function $g(t) \to 0$ (see Fig. S5b). When the barrier is low enough ($\approx 0.34\,kT$) to recover the ergodicity and mix two states, which occurs at time $\tau_f = 0.4\tau$, we start tilting. Later, the barrier is raised, which breaks the ergodicity. Finally, the potential tilt is removed. The tilt-controlling function $f(t)$ is shown in Fig S5c. Once the symmetric erasure is fully completed at $\frac{3}{4}\tau$, we expand the right well to its original size.

The protocol for the thermodynamically irreversible erasure never compresses the right well; rather, it erases the asymmetric memory directly. The function $m(t)$ is kept constant at all times (see Fig. S5a, red dotted line), while the rest remains the same as for the reversible protocol.

In Fig. S4, we plot potentials and their transformation at different stages in time, as well as the mean trajectories of particles for irreversible and reversible erasure. In both cases, when the bit is erased, a particle ends up in the right well; however, when time is reversed, the initial

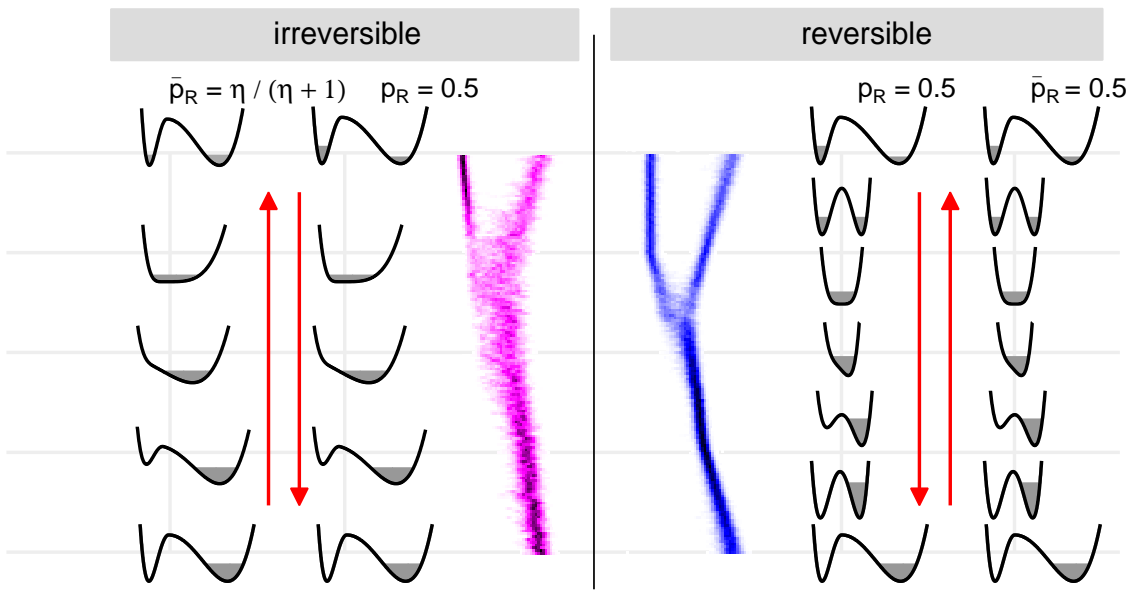

FIG. S4. Thermodynamically irreversible and reversible erasure protocols played forward and backward in time, along with typical mean trajectories. Red arrows indicate the direction of time. When time is reversed, the irreversible protocol does not return the system to the initial state, which had probabilities $p_L = p_R = 0.5$.

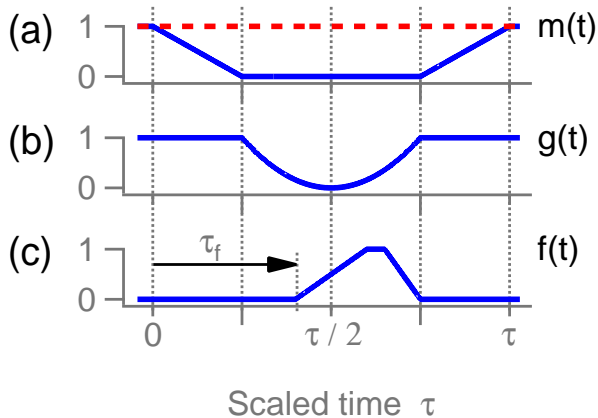

FIG. S5. Control functions for the erasure of an asymmetric bit. (a) Asymmetry control $m(t)$ for the thermodynamically reversible (blue line) and irreversible (red dotted line) protocols. (b) Control of the energy barrier height $g(t)$. (c) Tilt function $f(t)$. Tilt starts at $\tau_f = 0.4\tau$, when the barrier is sufficiently low and ergodicity is recovered.

probability of $\bar{p}_R = p_R = 0.5$ is recovered only for the thermodynamically reversible protocol. The probability to be in a well is schematically represented by the shaded area in a potential.

Note that in Fig. S5c, it is not necessary to wait to fully remove the barrier at $\tau_f = 0.5\tau$ to start tilting. Indeed, while Refs. [10, 16] took the conservative approach of completely lowering the barrier before starting to tilt, it is sufficient just to recover the ergodicity with a low barrier and start tilting at $\tau_f = 0.4\tau$. With this in mind, we start early, in order to make the erasure protocol 20% more efficient [17]. That is, work values will converge to their asymptotic values 20% faster. Since the data shown in the main article were collected over a two-week run, this shortened the runs by several days. Further decreasing $\tau_f$ would have led to unwanted trade-offs: If, for example, one starts tilting too early, when the barrier is too high and the ergodicity still broken, the symmetric erasure will be thermodynamically irreversible and the Landauer $kT \ln 2$ limit will not be reachable, even in the arbitrarily slow erasure limit.

## AN IRREVERSIBLE PROTOCOL

In the main article, we pointed out that Dillenschneider and Lutz (DL) [18] used a thermodynamically irreversible protocol to erase a symmetric bit. In their protocol, the tilting of the double-well potential starts too early, when the barrier is too high and the ergodicity still broken. This makes the DL erasure protocol thermodynamically irreversible, and it explains why all work values reported in that paper remain above the Landauer limit. But first, to explain the origin of thermodynamically irreversible symmetric erasure, we develop an analogy between the single-particle, double-well-potential systems and an ideal-gas experiment.

A system of two vessels and a valve contains an ideal gas at temperature $T$ (Fig. S6). Both vessels have the

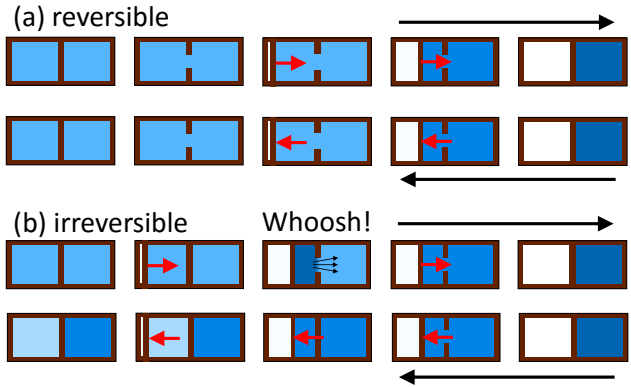

FIG. S6. Ideal gas in two identical separated vessels is analogous to a N-bit memory. Valve on divider can open to mix gases analogous to lowering the barrier in a double-well. Time direction is indicated by long black arrows. (a) Reversible protocol opens the valve first and then compresses to one vessel. When the protocol is played backward in time (below), it puts the gas in the same initial state. (b) Irreversible protocol starts compressing before the valve is opened. This creates a pressure difference between the two vessels, and free expansion occurs after the valve is opened. Playing the protocol backwards reveals the instability, as the gas does not return to the same initial state, as indicated by the shades of blue.

same volume and are filled with the same number of molecules. The erasure of memory is similar to the isothermal compression of an ideal gas to only one vessel. This compression can be done in a thermodynamically reversible way (a) if one opens the valve first, then compresses the gas to one vessel and closes the valve. When the protocol is time reversed, the gas goes through same states and ends up in the initial state. On the other hand, the compression can be done irreversibly (b) if one starts compressing first and then opens the valve. Since compressing one side creates a pressure difference, such a protocol is accompanied by free expansion [19], which is an irreversible step. The irreversibility can also be noticed when comparing the forward and time-reversed protocols. In the latter case, the initial state and the final state when time reversed are different.

Each vessel is analogous to one well of a double-well potential. Opening and closing the valve is similar to a continuous process of lowering and raising the barrier, and tilting the potential could be interpreted as creating a pressure difference between vessels. All transformations of the double-well potential are smooth and continuous, and the order in which those functions are applied determines the outcome of erasure, its efficiency, and its thermodynamic reversibility. In contrast to Fig. S6, where operations are done one at a time, some of the steps in the transformation of the double-well potential are done in parallel.

Figure S7 shows two possible ways to erase a symmetric one-bit memory, using 2d histograms of simulated trajectories. Part (a) describes the reversible protocol used here; part (b) corresponds to the kind of irreversible trajectory used in Ref. [18]. The parameters, described in detail below, are chosen to clearly distinguish the differing physical consequences of the two protocols. Qualitatively, in (a), ergodicity is established before tilting, and the protocol can be everywhere quasistatic. In (b), the tilt starts before ergodicity is established, leading to an inevitable irreversibility.

*Simulation details.* To understand the origin of irreversibility in (b), we consider a set of protocols parametrized by the tilt start time, $\tau_t$. Figure S8 shows the control functions $g(t)$ and $f(t)$ corresponding to Eq. S1.

The protocol starts from a double-well potential and a high barrier. It continuously lowers down the barrier (function $g(t)$ decreases to 0) and then raises it at the same rate immediately after. The tilt is introduced at different times, when the barrier has different heights. The function $f(t)$ initially equals zero. It then starts increasing at some time $\tau_t$, reaches its maximum, and then returns to 0.

We then simulate the erasure experiment for different tilt-start times $\tau_t$. Our simulation uses typical experimental parameters, and it also incorporates the effects of camera exposure, delay, and finite update times that are characteristic of feedback traps [9, 11]. We initially place particles around $+x_0$ for half of our measurements and around $-x_0$ for the other half. This ensures that the initial probability for a particle to be in the right well is $p_R = 0.5$. After the initial state is prepared, we let the particle equilibrate in the chosen well, and then we simulate the erasure experiment for a cycle time $\tau$. Later, we repeat the whole process for different values of $\tau$. For each $\tau$, we record many trajectories (see Fig. S7), and from them and the shape of applied potential, we find the mean work and also estimate work in the arbitrarily slow limit (see Fig. S9a). That procedure gives us the work in the arbitrarily slow limit for one value of the tilt-start time $\tau_t$.

We also do the entire protocol in reverse by taking the final state of erasure and applying control functions in reverse $g(\tau - t)$ and $f(\tau - t)$. For this reverse protocol, we estimate the probability for a particle to end up in the right $\bar{p}_R$ state. See Fig. S9b.

The mean work reported in Ref. [18] is $\langle W \rangle^* \approx 2.5\ kT \ln 2$. In principle, this excess over $\ln 2$ can come from two sources, the intrinsic irreversibility of the protocol and the finite cycle time of the protocol used in the simulations of Ref. [18]. Below, we will argue that the irreversibility is dominated by the intrinsic contribution.

In our simulation, $D = 0.23\ \mu\text{m}^2/\text{s}$, the distance between two potential minima is $2x_0 = 1.54\ \mu\text{m}$ and $E_b/kT = 13$, which correspond to our experimental val-

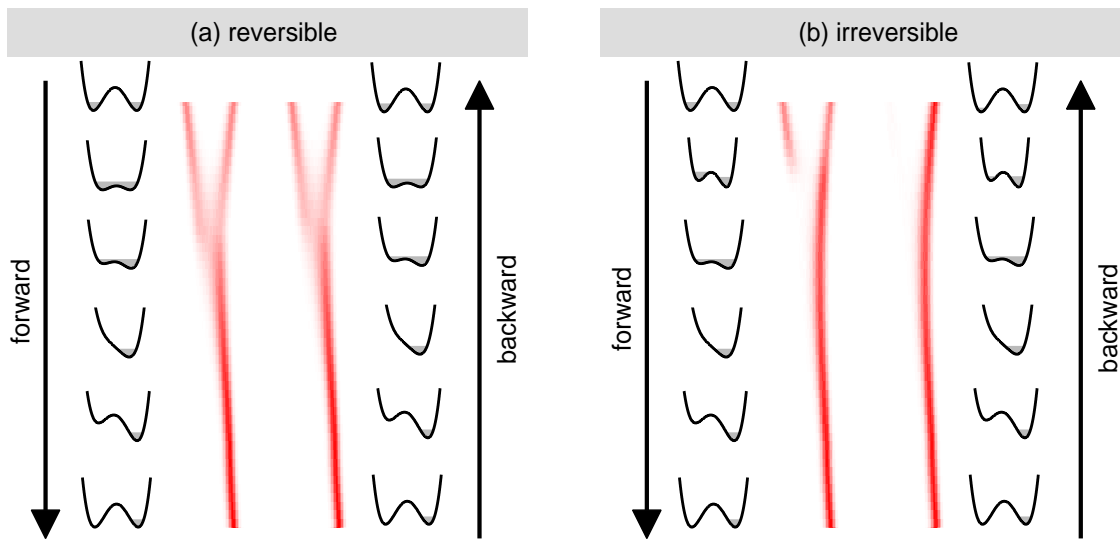

FIG. S7. Thermodynamically reversible and irreversible symmetric erasure protocols played forward and backward in time, along with simulated mean trajectories. (a) The barrier is lowered to allow for spontaneous hops, then we start tilting, raising the barrier and untilting. This protocol is reversible, and the probabilities for forward and backward trajectories are identical. (b) The irreversible protocol starts tilting and lowering the barrier at the same time. The system is not in global equilibrium when mixing of states occurs. One can clearly distinguish between forward and backward trajectories.

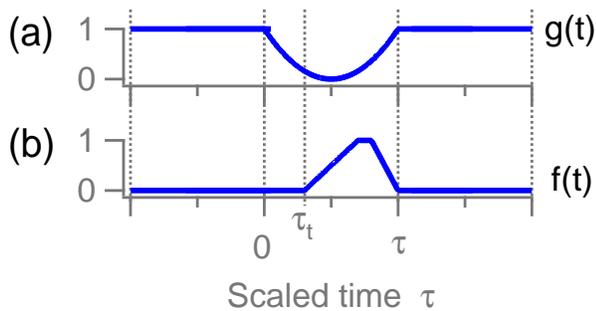

FIG. S8. Control functions for the symmetric erasure experiment. (a) Control of the energy barrier height $g(t)$. (c) Tilt function $f(t)$. Tilt starts at $\tau_f$ and controls if the protocol is going to be thermodynamically reversible.

ues. The scale time for erasure is then $\tau_0 = (2x_0)^2/D = 10.3$ s. In such a protocol, a dimensionless cycle time $\tau = 5$ suffices to reach $\ln 2$ within $0.16\ kT$.

By contrast, Ref. [18] uses $D^* = 10\ \mu\text{m}^2/\text{s}$, $2x_0^* \approx 20\ \mu\text{m}$, and $E_b^*/kT = 25$. Thus, $\tau_0^* = (2x_0^*)^2/D^* \approx 1$ s. Since the reported measurement time in Ref. [18] is 60 s, the scaled measurement cycle time is $\tau^* \approx 60$, which is deep into the asymptotic regime. Thus, we conclude that the remaining work excess is due to the intrinsic irreversibility of the protocol.

*Analysis.* The above conclusions are further supported by a series of simulations. When tilting starts too early ($\tau_t = 0$), the protocol requires more work than for tilts started later ($\tau_t = 0.3$). The difference is reflected in higher work values for all $\tau$ values in Fig. S9a. Further, we compare probabilities when time reversed $\bar{p}_R$, and they also deviate more from the initial $p_R = 0.5$ for $\tau_t = 0$. Generally, if one of the initial probabilities is not recovered when time reversed $\bar{p}_i \neq p_i$, the protocol must be irreversible; however, even if all initial probabilities are recovered $\bar{p}_i = p_i$ (for all $i$), the protocol is not necessarily reversible. Since the first outcome occurs here, we conclude that the protocol for $\tau_t = 0$ must be thermodynamically irreversible.

We repeat the simulation for several different values of $\tau_t$, ranging from 0 to 0.45, and explore which $\tau_t$ remains thermodynamically reversible. Figure S10 shows estimated work and probabilities in an arbitrarily slow limit for different starting times for tilt, parametrized by $\tau_t$. The increase in minimal work to erase occurs for $\tau_t \lesssim 0.2$ (Fig. S10b); therefore, for the symmetric-erasure component of the asymmetric-erasure protocol, we use $\tau_t = 0.3$.

To test that this increase in work is in fact caused by thermodynamic irreversibility, we start again from Eq. S10, which is appropriate for our case, because there is only one potentially irreversible step, which is lowering the barrier. If there were several irreversible steps in the protocol, one would include the relative entropy $D_{KL}(p_i||\bar{p}_i)$ for each irreversible step.

For symmetric potentials, according to Eq. S11 $\Delta F = 0$. The information erased is $H = 1$ bit $= \ln(2)$ nat. Since our system has only two macrostates, we simplify indices $i$ to left ($i = 1 = L$) and right ($i = 2 = R$). Initially, the probabilities to be in these states are set to

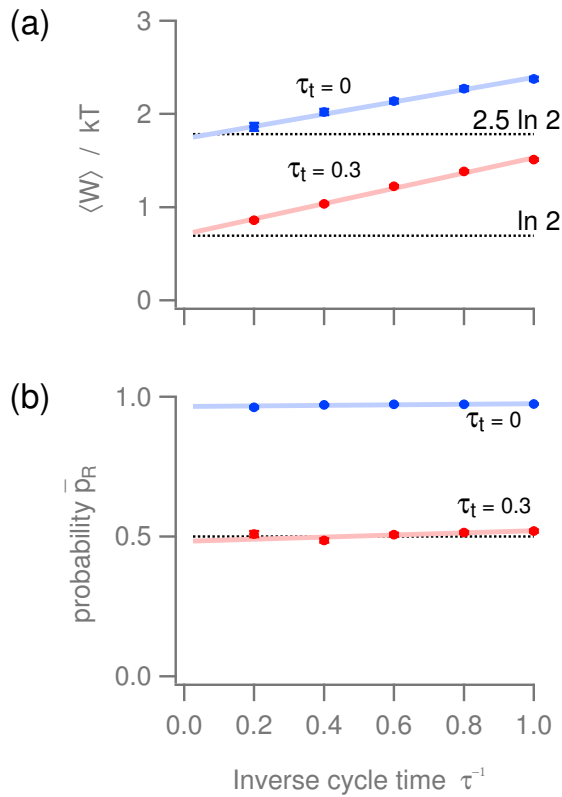

FIG. S9. Simulation exploring work and reversibility of a symmetric erasure protocol. (a) Work to erase a symmetric memory for two different protocols, where tilting starts at $\tau_f = 0$ and $\tau_f = 0.3$. The protocol that starts to tilt earlier, at $\tau_f = 0$, requires more work. The Landauer limit ($\ln 2$) and mean work in DL ($\approx 2.5 \ln 2$) are shown using dotted lines. (b) Probability of ending up in the right well after time is reversed.

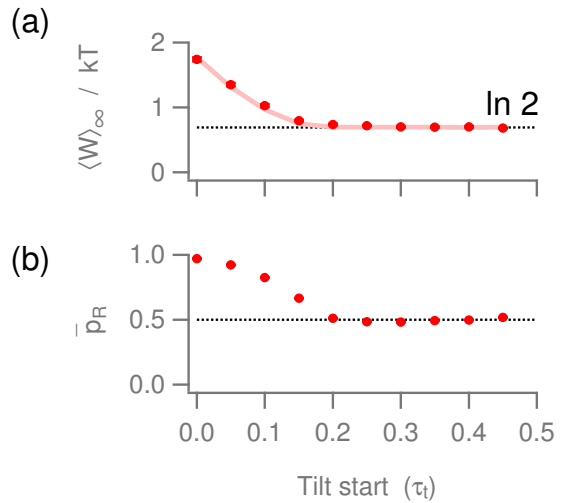

FIG. S10. Simulation testing work and the efficiency of the erasure protocol. (a) Work depends on where the tilting starts. If tilting starts too early, $\tau_t \lesssim 0.2$, more work is required to erase memory due to thermodynamic irresistibility. (b) Probability of returning to the right state when time revered depends on when tilting of the potential start. For early tilting, $\tau_f \lesssim 0.2$, this probability is different than the initial probability $\bar{p}_R \neq p_R$.

$p_L = p_R = 0.5$, but when time is reversed we "measure" $\bar{p}_R(\tau_t)$ and obtain $\bar{p}_L(\tau_t) = 1 - \bar{p}_R(\tau_t)$. Next, we write the work to erase a one-bit memory in this type of irreversible protocol:

$$\langle W \rangle /kT \geq -\tfrac{1}{2} \ln \bar{p}_R - \tfrac{1}{2} \ln(1 - \bar{p}_R) \quad \text{(S16)}$$

where $\bar{p}_R$ depends on the details of the protocol and puts lower bound on work in such protocol. This inequality becomes an equality in the arbitrarily slow limit. Thus,

$$\langle W \rangle_\infty /kT = -\tfrac{1}{2} \ln \bar{p}_R - \tfrac{1}{2} \ln(1 - \bar{p}_R) \quad \text{(S17)}$$

We plot this prediction as a solid red line in Fig. S10a, where, for $\bar{p}_R$, we used estimated values from (b). When $\bar{p}_R$ is 0.5, the work to fully erase a one-bit memory in an arbitrarily slow limit is $kT \ln 2$; in other cases, it is more. The work in Eq. S17 diverges to infinity for $\bar{p}_R = 0$ or $\bar{p}_R = 1$. We keep the tilt parameter $A$ (Eq. S1) fixed for all simulations. It is straightforward to see that changing $A$ will change $\bar{p}_R$ and therefore the total work in the forward direction.

Finally, we qualitatively compare work distributions for $\tau_t = 0$ (corresponding to the Dillenschneider-Lutz protocol and $\tau_t = 0.3$ (corresponding to our protocol, where the duration of the erasure protocol is $\tau = 5$ ( $\tau^{-1} = 0.2$).

For the irreversible protocol in Fig. S11a, the work distribution is bimodal, with each peak corresponding to a different particle trajectory. One peak corresponds to particles that start in the left well, and the other corresponds to particles that start in the right well. In the irreversible protocol, particles do not lose correlation with their initial state.

For the reversible protocol in Fig. S11b, the distribution is Gaussian, and from the measured work value, one cannot say which well the particle came from. The work to lower the barrier, mix states, and restore the ergodicity is the same for particles starting in either well, because the potential is symmetric during such processes. Once ergodicity is restored, particles can cross the low barrier and freely explore both sides of the potential. At that point, it loses correlation with its initial state, so that the work to tilt and raise the barrier becomes independent of the particle's origin.

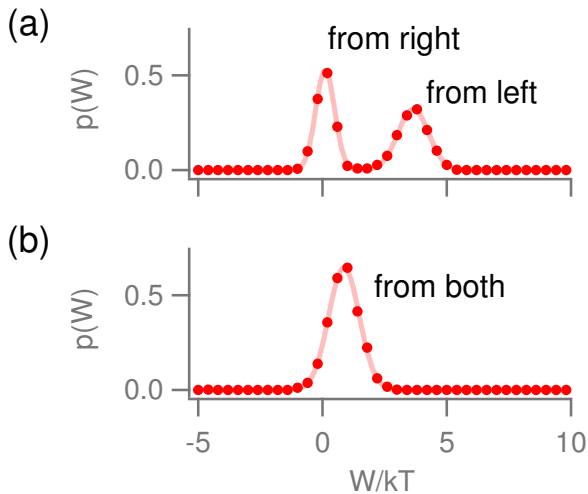

FIG. S11. Simulated work distribution for thermodynamically irreversible and reversible protocols for cycle time $\tau = 5$. (a) For the irreversible protocol ($\tau_t = 0$), the work distribution is bimodal, with each peak corresponding to one of the initial states. The average work to erase the 0 state differs from that needed to erase the 1 state. (b) For our reversible protocol ($\tau_t = 0.3$), the work distributions for a particle starting in either state (well) are indistinguishable. Thus, the average work to erase the 0 and 1 states is identical.

[1] M. Gavrilov and J. Bechhoefer, EPL (Europhysics Letters) **114**, 50002 (2016).
[2] A. Weigel, A. Sebesta, and P. Kukura, ACS Photonics **1**, 848 (2014).
[3] M. Gavrilov, J. Koloczek, and J. Bechhoefer, in *Novel Techniques in Microscopy* (Opt. Soc. Am., 2015) p. JT3A. 4.
[4] A. J. Berglund, M. D. McMahon, J. J. McClelland, and J. A. Liddle, Opt. Express **16**, 14064 (2008).
[5] M. Gavrilov, Y. Jun, and J. Bechhoefer, Proc. SPIE **8810** (2013).
[6] K. Sekimoto, J. Phys. Soc. Jap. **66**, 1234 (1997).
[7] K. Sekimoto, *Stochastic Energetics* (Springer, 2010).
[8] J. Happel and H. Brenner, *Low Reynolds Number Hydrodynamics: With Special Applications to Particulate Media* (Martinus Nijhoff, 1983).
[9] Y. Jun and J. Bechhoefer, Phys. Rev. E **86**, 061106 (2012).
[10] Y. Jun, M. Gavrilov, and J. Bechhoefer, Phys. Rev. Lett. **113**, 190601 (2014).
[11] M. Gavrilov, Y. Jun, and J. Bechhoefer, Rev. Sci. Instrum. **85**, 095102 (2014).
[12] R. Kawai, J. M. R. Parrondo, and C. Van den Broeck, Phys. Rev. Lett. **98**, 080602 (2007).
[13] J. M. R. Parrondo, C. Van den Broeck, and R. Kawai, New J. Phys. **11**, 073008 (2009).
[14] É. Roldán, I. A. Martínez, J. M. R. Parrondo, and D. Petrov, Nature Phys. **10**, 457 (2014).
[15] T. Sagawa and M. Ueda, *Information Thermodynamics: Maxwell's Demon in Nonequilibrium Dynamics*, edited by R. Klages, W. Just, and C. Jarzynski (Wiley-VCH, Weinheim, 2013, 2013).
[16] D. Chiuchiú, EPL (Europhysics Letters) **109**, 30002 (2015).
[17] P. R. Zulkowski and M. R. DeWeese, Phys. Rev. E **89**, 052140 (2014).
[18] R. Dillenschneider and E. Lutz, Phys. Rev. Lett. **102**, 210601 (2009).
[19] H. B. Callen, *Thermodynamics and an Introduction to Thermostatistics*, 2nd ed. (Wiley, 1985).



\end{document}